\documentclass[10pt]{IEEEtran}
\usepackage{cite}
\usepackage{amsmath,amssymb,amsfonts}

\usepackage{amsthm} 

\usepackage{algorithmic}
\usepackage{graphicx}

\usepackage{epstopdf}
\usepackage{colortbl}
\usepackage{diagbox}
\usepackage{mathtools}
\usepackage{empheq}
\newcommand*\widefbox[1]{\fbox{\hspace{2em}#1\hspace{2em}}}
\usepackage{tikz}
\usepackage{pifont}
\newcommand{\cmark}{\ding{51}}%
\newcommand{\xmark}{\ding{55}}%

\DeclareMathOperator*{\argmin}{arg\ min}

\newtheorem{mydef}{Definition}
\newtheorem{mylem}{Lemma}
\newtheorem{mypbm}{Problem}

\newtheorem{myrem}{Remark}

\newtheorem{mycor}{Corollary}
\newtheorem{myprs}{Proposition}

\usepackage[linesnumbered,boxed,commentsnumbered,ruled,vlined,longend]{algorithm2e}

\SetAlgorithmName{Procedure}

\usepackage{comment}

\usepackage[colorlinks = true,
linkcolor = blue,
urlcolor  = blue,
citecolor = blue,
anchorcolor = blue]{hyperref}
\usepackage{comment}

\title{On the Constrained CAV Platoon Control Problem}

\author{MirSaleh Bahavarnia$^\dagger$, Junyi Ji$^{\dagger\P}$  Ahmad F. Taha$^\dagger$, and Daniel B. Work$^{\dagger\P}$ 
	\thanks{$^\dagger$The authors are with the Department of Civil and Environmental Engineering, Vanderbilt University, 2201 West End Avenue, Nashville, TN 37235, USA. $^\P$Institute for Software Integrated Systems. Emails: mirsaleh.bahavarnia@vanderbilt.edu, junyi.ji@vanderbilt.edu, ahmad.taha@vanderbilt.edu, dan.work@vanderbilt.edu.}
 \thanks{This work is supported by the National Science Foundation, United States under grants ECCS 2151571, CMMI 2152450 and 2152928, and CNS 2135579.}
}

\begin{document}
	


\maketitle

\begin{abstract}
The main objective of the connected and automated vehicle (CAV) platoon control problem is to regulate CAVs' position while ensuring stability and accounting for vehicle dynamics. Although this problem has been studied in the literature, existing research has some limitations. This paper presents two new theoretical results that address these limitations: \textit{(i)} the synthesis of unrealistic high-gain control parameters due to the lack of a systematic way to incorporate the lower and upper bounds on the control parameters, and \textit{(ii)} the performance sensitivity to the communication delay due to inaccurate Taylor series approximation. To be more precise, taking advantage of the well-known Padé approximation, this paper proposes a constrained CAV platoon controller synthesis that \textit{(i)} systematically incorporates the lower and upper bounds on the control parameters, and \textit{(ii)} significantly improves the performance sensitivity to the communication delay. The effectiveness of the presented results is verified through conducting extensive numerical simulations. The proposed controller effectively attenuates the \textit{stop-and-go disturbance}---a single cycle of deceleration followed by acceleration---amplification throughout the mixed platoon (consisting of CAVs and human-driven vehicles). Modern transportation systems will benefit from the proposed CAV controls in terms of effective disturbance attenuation as it will potentially reduce collisions.
\end{abstract}


\section{Introduction and Paper Contributions}\label{sec:Intro}

\IEEEPARstart{T}{raffic} oscillations, also known as \textit{stop-and-go disturbances}, in congested traffic have attained an increasing amount of attention in recent years (see \cite{zhou2020stabilizing} and the relevant references therein). The term \textit{stop-and-go disturbance} refers to a single cycle of deceleration followed by acceleration. Various approaches have been developed to mostly control human-driven vehicles. One promising approach is via variable speed limit (VSL) control \cite{popov2008distributed,hegyi2009expected}. Another one utilizes the emerging connected and automated vehicle (CAV) technology which facilitates the effective control of CAVs leading to significant enhancements in traffic flow capacity and stability \cite{chen2017towards,piacentini2019highway,zhou2020stabilizing,vishnoi2023cav}. Specifically, the main objective of the \textit{CAV platoon control problem} is to regulate CAVs' position while ensuring stability and accounting for vehicle longitudinal dynamics \cite{zhou2020stabilizing}. 

CAV control strategies can mainly be classified into two controller types: \textit{(i)} model predictive control (MPC) \cite{wang2014rolling,gong2016constrained,ma2017parsimonious,zhou2017rolling,piacentini2019highway}, and \textit{(ii)} linear controller \cite{van2006impact,naus2010string,morbidi2013decentralized,zhou2020stabilizing,vishnoi2023cav}. Tab. \ref{tab0:my_label} reflects the pros and cons corresponding to such controller types in terms of their ability to incorporate two crucial features: \textit{(i)} \textit{local stability} and \textit{string stability} (two main stability constraints to be thoroughly detailed in Section~\ref{SecP}), and \textit{(ii)} explicit constraints (e.g., lower and upper bounds on the control parameters). Inspired by Tab. \ref{tab0:my_label}, in the current study, we choose linear controller as the controller type and propose a systematic procedure to effectively incorporate both \textit{(i)} local stability and string stability (as stability constraints) and \textit{(ii)} \textit{box constraints} (i.e., the lower and upper bounds) on the control parameters (as a main subclass of explicit constraints).

\begin{table}[t]
\caption{The general schematic structure of feature incorporation for two main controller types in the CAV control strategies: \textit{(i)} MPC, and \textit{(ii)} Linear Controller.}
    \centering
    \begin{tabular}{|c|c|c|}
    \hline
        \diagbox{Feature}{Controller Type} & MPC & Linear Controller \\
        \hline
        Local Stability and String Stability & \xmark & \cmark \\
        \hline
        Explicit Constraints & \cmark & \xmark \\
        \hline
    \end{tabular}
    \label{tab0:my_label}
\end{table}

In CAV-centered studies, vehicle platooning \cite{larsson2015vehicle} emerges as a pivotal application with significant potential in both the immediate and distant future. This technology, crucial for enhancing road safety and efficiency, heavily relies on the efficacy of vehicle-to-everything (V$2$X) communication systems. Research, including findings from \cite{molina2017lte}, underscores that both dedicated short-range communications and cellular vehicle-to-everything technologies can competently support safety applications necessitating end-to-end latency around $100$ milliseconds, provided that vehicle density remains within manageable limits. However, a critical challenge arises as traffic density escalates, leading to a marked increase in V$2$X communication delay due to the communication channel congestion, as noted in several studies \cite{dey2016vehicle,ahmad2019v2v,naik2019ieee}. This surge in latency, particularly in congested scenarios, is a matter of concern. 

As highlighted in \cite{naik2019ieee}, effective vehicle platooning must accommodate maximum latencies ranging from $10$ to $500$ milliseconds. This necessitates the development of advanced control systems capable of adapting to these varying delay conditions, thereby ensuring the reliability and safety of vehicle platooning in diverse traffic environments. This research aims to address the escalating delays in high-density traffic scenarios, emphasizing the need for advanced controller synthesis that can effectively manage these challenges in vehicle platooning applications.

\noindent {\bf Paper Objectives and Contributions.} The main objective of the CAV platoon control problem is to regulate CAVs' position while ensuring stability and accounting for vehicle longitudinal dynamics \cite{zhou2020stabilizing}. Although such a problem has been studied thoroughly in the literature, the existing research work \cite{zhou2020stabilizing} has some limitations. Two main limitations include \textit{(i)} the synthesis of unrealistic high-gain control parameters due to the lack of a systematic way to incorporate the lower and upper bounds on the control parameters, and \textit{(ii)} the performance sensitivity to the communication delay due to inaccurate Taylor series approximation. The paper's contributions can be summarized as follows:
\begin{itemize}
    \item To effectively address such limitations, taking advantage of the well-known Padé approximation, this paper proposes a constrained CAV platoon controller synthesis that \textit{(i)} systematically incorporates the lower and upper bounds on the control parameters, and \textit{(ii)} significantly improves the performance sensitivity to the communication delay. 
    \item Given box constraints on the control parameters, we first parameterize the locally stabilizing feedback and feedforward gains. Deriving a necessary condition to ensure the string stability criterion additionally, we then obtain more representative parameterized locally stabilizing feedback and feedforward gains. Utilizing the Padé approximation, we more accurately ensure the string stability criterion in the presence of communication delay compared to the widely utilized Taylor series approximation. Furthermore, in the case of a sufficiently small communication delay, the Padé approximation attains a better near-optimality and in the case of a large communication delay, it successfully obtains a near-optimal CAV platoon controller synthesis while the widely utilized Taylor series approximation counterpart becomes unusable.
    \item Considering the mixed vehicular platoon scenario and minimizing the $\mathcal{H}_{\infty}$ norm of the Padé approximated transfer function over an interval defined by predominant acceleration frequency boundaries of human-driven vehicles, we solve for a near-optimal constrained CAV platoon controller synthesis via nonlinear optimization tools. The effectiveness of the presented results is verified through conducting extensive numerical simulations. The proposed CAV platoon controller synthesis can effectively attenuate the stop-and-go disturbance amplification throughout the mixed vehicular platoon. Modern transportation systems will potentially benefit from the proposed CAV control strategy in terms of effective stop-and-go disturbance attenuation as it will potentially reduce collisions.
\end{itemize}

The remainder of the paper is structured as follows: Section \ref{sec:ProFor} formally presents the preliminaries and states the problem to be solved. Section \ref{MRs} contains the main results followed by numerical simulations illustrated by Section \ref{secNS}. Finally, Section \ref{Con} ends the paper with a few concluding remarks.  


\noindent {\bf Notation.} The uppercase and lowercase letters denote the matrices and vectors, respectively. We represent the supremum and maximum by $\sup$ and $\max$, respectively. The set of $n$-dimensional real-valued vectors is symbolized by $\mathbb{R}^n$. For a vector $v \in \mathbb{R}^n$, we denote its $\ell_{\infty}$ norm (i.e., $\max_{i} |v_i|$) by $\|v\|_{\ell_{\infty}}$. We represent the vector of all ones with $\mathbf{1}$. To symbolize the imaginary unit, we use $j = \sqrt{-1}$. For a square matrix $M$, we denote the set of its eigenvalues with $\lambda(M)$ and show its spectral abscissa (i.e., the maximum real part of its eigenvalues) by $\mathbf{sa}(M)$. A square matrix $M$ is said to be Hurwitz if $\mathbf{sa}(M) < 0$ holds. The time derivative of a signal $\chi(t)$ is represented by $\dot{\chi}(t)$. For a function $f(\omega)$, the first derivative and the second derivative with respect to $\omega$ are respectively denoted by $\frac{d f(\omega)}{d \omega}$ and $\frac{d^2 f(\omega)}{d \omega^2}$. Also, for brevity, we alternatively use $f'(\omega)$ and $f''(\omega)$ to refer to the first and second derivatives, respectively. To represent the limit of a function $f(\omega)$ as $\omega$ tends to $a$, we utilize the notation $\lim_{\omega \to a} f(\omega)$. 

\section{Preliminaries and Problem Statement} \label{sec:ProFor}

This section is divided into two main parts: \textit{(a)} preliminaries in Section~\ref{SecP}, and \textit{(b)} problem statement in Section~\ref{SecProb}. Preliminaries consist of \textit{(i)} the state-space representation, \textit{(ii)} the local stability, and \textit{(iii)} the string stability of the CAV system. 

Built upon the preliminaries, we next state a control problem to propose a constrained CAV platoon controller synthesis subject to the lower and upper bounds on the control parameters to prevent the synthesis of unrealistic high-gain control parameters. Furthermore, we will observe that communication delay can effectively be handled by utilizing a more accurate approximation for the delay-dependent (exponential) term.

According to the travel direction, we consider a vehicular platoon consisting of $N_{\mathrm{vp}}$ vehicles with vehicle $1$ and vehicle $N_{\mathrm{vp}}$ as the most preceding (the first) and the most following (the last) vehicles, respectively.

\subsection{Preliminaries} \label{SecP}

\subsubsection{State-space representation}
Let us consider the state-space representation of the CAV system $i$ as follows \cite{swaroop1994comparision,yi2001vehicle,zhou2020stabilizing}:
\begin{align} \label{CAVSys}
    \dot{\chi}_i(t) &= A_i {\chi}_i(t) + B_i u_i(t) + D a_{i-1}(t), 
\end{align}
with
\begin{align*}
    A_i &= \begin{bmatrix}
        0 & 1 & -\tau^{\ast}_i\\
        0 & 0 & -1\\
        0 & 0 & -\frac{1}{T_{i}}
    \end{bmatrix},B_i = \begin{bmatrix}
        0\\0\\\frac{K_{i}}{T_{i}}
    \end{bmatrix},D = \begin{bmatrix}
        0\\1\\0
    \end{bmatrix},
\end{align*}
where ${\chi}_{i}(t) = \begin{bmatrix}
    \sigma_i(t) & \Delta v_i(t) & a_i(t)
\end{bmatrix}^\top$, $\sigma_i(t)$, $\Delta v_i(t) := v_{i-1}(t) - v_i(t)$, $a_i(t)$, and $u_i(t)$ denote the state vector, the deviation from equilibrium spacing, the speed difference with the preceding vehicle (i.e., vehicle $i-1$), the realized acceleration, and the control input, respectively. Moreover, $\tau^{\ast}_i$, $T_{i}$, and $K_{i}$ represent the predefined constant time gap, time-lag for vehicle $i$ to realize the acceleration, and ratio of demanded acceleration that can be realized, respectively. We can find an equilibrium state $\chi_{i,e}(t)$ by setting $\chi_{i}(t) = \chi_{i,e}(t) = \begin{bmatrix}
    0 & 0 & 0
\end{bmatrix}^\top$. Also, note that $a_{i-1}(t)$ is treated as an external disturbance for CAV system $i$ because it is not controllable by CAV system $i$.

As highlighted in \cite{zhou2020stabilizing}, to bypass multiple delay accumulation, we utilize the following standard decentralized linear control strategy:
\begin{align} \label{DLC}
    u_i(t) &= \begin{bmatrix}
        k_{si} & k_{vi} & k_{ai}
    \end{bmatrix} {\chi}_i(t) + k_{fi} a_{i-1}(t-\theta),
\end{align}
where $k_{si}$, $k_{vi}$, and $k_{ai}$ respectively represent the feedback gains associated with the derivation from equilibrium spacing $\sigma_i$, the speed difference $\Delta v_i$, and the acceleration $a_i$. Moreover, $k_{fi}$ and $\theta$ denote the feedforward gain and V$2$V/V$2$I communication delay, respectively. For the sake of brevity, we alternatively utilize $k = \begin{bmatrix}
        k_1 & k_2 & k_3 & k_4
    \end{bmatrix}$ to denote $k = \begin{bmatrix}
        k_{si} & k_{vi} & k_{ai} & k_{fi}
    \end{bmatrix}$.
Similarly, we simply denote $A_i$, $B_i$, $K_i$, $T_i$, and $\tau_i^{\ast}$ by $A$, $B$, $K$, $T$, and $\tau$, respectively.

The main objective of the linear control strategy \eqref{DLC} is to regulate CAVs' position while ensuring stability and accounting for vehicle longitudinal dynamics \cite{zhou2020stabilizing}. The linear control strategy \eqref{DLC} has widely been utilized in the literature \cite{van2006impact,morbidi2013decentralized,zhou2020stabilizing}. As highlighted in Tab. \ref{tab0:my_label}, one important advantage of the linear control strategy \eqref{DLC} over the MPC control strategy is that it enables us with stability guarantees \cite{zhou2020stabilizing}. Furthermore, later on, we show that by taking advantage of the parameterization of stability regions of the CAV system in \eqref{CAVSys} governed by the control strategy in \eqref{DLC} subject to imposed box constraints on the control parameters, we effectively embed the constraints and minimize the $\mathcal{H}_{\infty}$-based objective function similar to the MPC control strategy.

\subsubsection{Local stability}

Local stability (also known as internal stability) means that a disturbance (deviation from an equilibrium point) can locally be resolved by a system. In the context of a CAV system, it means that deviation from an equilibrium spacing, speed difference, and acceleration can locally be resolved in a vehicle \cite{zhou2020stabilizing}.    

\begin{mydef}[Local stability \cite{li2018robust,zhou2020stabilizing}]
    A CAV platoon following a linear control strategy (e.g., the control strategy in \eqref{DLC}) is called locally stable with respect to an equilibrium point $\chi_{e}$ if and only if $\mathbf{sa}(A+Bk) < 0$ holds ($A+Bk$ is Hurwitz).
\end{mydef}

As \href{https://www.sciencedirect.com/science/article/pii/S0191261518311482?casa_token=NzZJbHpjaSwAAAAA:UPUagh0REEv0nvScPTw_PEmOfgaPnfwXx1TWtitr4Hg7rL2FlOfbiWUy3DdbkHuXWjxaGkC2}{Proposition $1$} in \cite{zhou2020stabilizing} states, the CAV system in \eqref{CAVSys} governed by the control strategy in \eqref{DLC} is locally stable if the following inequalities are satisfied:
\begin{subequations} \label{LSC}
    \begin{align}
        & k_1 > 0,\\
        & \tau k_1+k_2 > 0,\\
        & \frac{1}{K}-k_3 > 0,\\
        & \bigg(\frac{1}{K}-k_3\bigg)(\tau k_1+k_2)-\frac{T}{K}k_1 >0.
    \end{align}    
\end{subequations}It is noteworthy that inequalities \eqref{LSC} are derived by applying the Hurwitz criterion \cite{parks1962new} to the cubic polynomial $\det (sI-(A+Bk)) = \frac{T}{K}s^3 + (\frac{1}{K}-k_3)s^2 + (\tau k_1 + k_2)s + k_1$. Applying the Hurwitz criterion \cite{parks1962new}, ensures that all the elements of $\lambda(A+Bk)$ lie on the LHS of the imaginary axis, i.e., $\mathbf{sa}(A+Bk) < 0$ holds ($A+Bk$ is Hurwitz).

\subsubsection{String stability}
Strict string stability (also known as $\mathcal{H}_2$ norm string stability) means that the magnitude of a disturbance is not amplified for each leader-follower pair through a vehicular string \cite{zhou2020stabilizing}. 

\begin{mydef}[Strict string stability \cite{naus2010string,zhou2020stabilizing}] A CAV vehicular string is strict string stable if and only if 
\begin{align*}
    & \frac{\|a_i(s)\|_2}{\|a_{i-1}(s)\|_2} \le 1,~\forall i \in \mathcal{I}_{\mathrm{CAV}}, 
\end{align*}
hold where $\mathcal{I}_{\mathrm{CAV}}$ denotes the set of CAVs and $\|a_i(s)\|_2$ represents the $\mathcal{H}_2$ norm of $a_i(s)$ which is defined as $\|a_i(s)\|_2 := \sqrt{\int_0^{\infty} |a_i(j \omega)|^2 d \omega} $. ($\|a_{i-1}(s)\|_2$ follows the similar notation/definition.)    
\end{mydef}

As expressed by \cite{zhou2020stabilizing}, applying the following Cauchy inequality \cite{naus2010string}: 
\begin{align*}
    \frac{\|a_i(s)\|_2}{\|a_{i-1}(s)\|_2} \le \bigg \|\frac{a_i(s)}{a_{i-1}(s)} \bigg \|_{\infty},
\end{align*}
a sufficient condition to guarantee the string stability can be derived as
\begin{align} \label{SCon}
\|F_i(s)\|_{\infty} &:= \underset{\omega > 0}{\sup}~|F_i(j\omega)| \le 1,
\end{align}
where $F_i(s) = \frac{a_i(s)}{a_{i-1}(s)}$ and $\|F_i(s)\|_{\infty}$ denote the transfer function capturing disturbance propagation through the vehicular string in the frequency domain and its $\mathcal{H}_{\infty}$ norm (i.e., the upper bound of the disturbance propagation ratio through the vehicular string), respectively. As a reminder, $a_{i-1}(t)$ (equivalently, $a_{i-1}(s)$ in frequency domain) is treated as an external disturbance which facilitates the $\mathcal{H}_{\infty}$ norm applications in this context.

By taking the Laplace transform of the closed-loop system consisting of \eqref{CAVSys} and \eqref{DLC}, one can obtain the following expression for $F_i(s)$:
\begin{align} \label{TF}
    & F_i(s) = \frac{K(k_4 s^2e^{-\theta s}+k_2 s+ k_1)}{c_3s^3+c_2s^2+c_1s + c_0},\\
    & c_3 = T,~c_2 = -K k_3+1,~c_1 = K(\tau k_1 +k_2),~
    c_0 = K k_1. \notag
\end{align}
Also, it can be verified that $|F_i(j\omega)|$ can be computed as
    \begin{align} \label{Fabs}
        &|F_i(j\omega)| = \sqrt{\frac{N(\omega)}{D(\omega)}},\\
        & N(\omega) = n_4 \omega^4 + (n_2+ g_{\theta}(\omega)) \omega^2 + n_0,\notag\\
        & D(\omega) = d_6 \omega^6 + d_4 \omega^4 + d_2 \omega^2 + d_0,\notag\\
        &n_4 = K^2 k_4^2,~
        n_2 = K^2 k_2^2,~
        n_0 = K^2k_1^2,\notag\\
        &g_{\theta}(\omega) = 2 K^2 k_4(-k_1 \cos (\theta \omega) + k_2 \omega \sin (\theta \omega)),\notag\\
        &d_6 = T^2,~
        d_4 = -2 K T(\tau k_1+k_2)+ (K k_3-1)^2,\notag\\
        &d_2 = K^2 (\tau k_1 +k_2)^2 + 2K k_1(Kk_3-1),~d_0 = K^2 k_1^2. \notag
    \end{align}
It is noteworthy that the exponential term associated with the communication delay $\theta$ in $F_i(s)$ in \eqref{TF}, i.e., $e^{-\theta s}$, adds more complexity to the $\mathcal{H}_{\infty}$ controller synthesis. In other words, we need to utilize approximation techniques to efficiently compute the approximate value of $\|F_i(s)\|_{\infty}$. To that end, there exist various approximations. For instance, the authors in \cite{zhou2020stabilizing}, built upon a sufficient condition (i.e., utilizing the Taylor series approximations of $\cos(\theta \omega) \approx 1-\frac{\theta^2 \omega^2}{2}$ and $\sin(\theta \omega) \approx \theta \omega - \frac{\theta^3 \omega^3}{6}$ when $\theta$ is sufficiently small) and imposing the sufficient condition \eqref{SCon}, derive a set of inequalities presented by \href{https://www.sciencedirect.com/science/article/pii/S0191261518311482?casa_token=NzZJbHpjaSwAAAAA:UPUagh0REEv0nvScPTw_PEmOfgaPnfwXx1TWtitr4Hg7rL2FlOfbiWUy3DdbkHuXWjxaGkC2}{Proposition $2$} therein as
    \begin{subequations} \label{SSC}
        \begin{align}
        &~T^2 + \frac{K^2k_4k_2\theta^3}{3} \ge 0,\label{SSC1} \\
        &-2KT(\tau k_1 + k_2)+(Kk_3-1)^2 \notag \\& -K^2k_4 (k_1\theta^2+2k_2\theta+k_4) \ge 0,\label{SSC2} \\
        &~2Kk_1 \bigg( K \bigg(k_4 + k_3 + \tau k_2 + \frac{\tau^2 k_1}{2}\bigg)-1 \bigg) \ge 0, \label{SSC3}
    \end{align}
    \end{subequations} to guarantee the string stability of the CAV system in \eqref{CAVSys} governed by the control strategy in \eqref{DLC} when $\theta$ is sufficiently small. The proof of \href{https://www.sciencedirect.com/science/article/pii/S0191261518311482?casa_token=NzZJbHpjaSwAAAAA:UPUagh0REEv0nvScPTw_PEmOfgaPnfwXx1TWtitr4Hg7rL2FlOfbiWUy3DdbkHuXWjxaGkC2}{Proposition $2$} in \cite{zhou2020stabilizing} relies on a strong condition. In other words, to impose the non-negativeness of the corresponding quartic polynomial, namely, $p\omega^4 + q \omega^2 + r \ge 0$ for which $p$, $q$, and $r$ respectively denote the expressions on the LHS of \eqref{SSC1}--\eqref{SSC3}, they require all coefficients $p$, $q$, and $r$ to be non-negative as expressed by inequalities \eqref{SSC1}--\eqref{SSC3}. However, the feasibility set associated with the string stability can be expanded by additionally including the following case:
\begin{align}
    p \ge 0,~q < 0,~r \ge 0,~ q^2-4pr \le 0,
\end{align}
which is missing in \cite{zhou2020stabilizing}. Such an inclusion helps us to search a broader space and potentially find a better near-optimal solution which was not achievable in \cite{zhou2020stabilizing} previously.

\subsection{Problem statement} \label{SecProb}

It is noteworthy that the ultimate goal is to apply the CAV platoon controller synthesis results to the mixed vehicular platoon. Due to the string unstable behavior of human-driven vehicles, we have no direct control over them. However, applying the (fully) CAV platoon controller synthesis results to the mixed vehicular platoon can effectively attenuate the stop-and-go disturbance amplification throughout the mixed vehicular platoon. In the case of the mixed vehicular platoon, the frequency of human-driven vehicle acceleration throughout traffic oscillations is bounded and typically shows a predominant range \cite{thiemann2008estimating,zhou2020stabilizing}. Then, for the sake of practicality, we focus on minimizing $\|F_i(s)\|_{\infty}$ over an interval defined by predominant acceleration frequency boundaries of human-driven vehicles (see \cite{zhou2020stabilizing} for more details).

Given the transfer function $F_i(s) = \frac{a_i(s)}{a_{i-1}(s)}$ and the predominant acceleration frequency boundaries of human-driven vehicles, namely, $\omega_2 > \omega_1 > 0$, let us define its $\mathcal{H}_{\infty}$ norm over $\omega \in [\omega_1,\omega_2]$, namely, $\mathcal{H}_{\infty}^{[\omega_1,\omega_2]}$, as \cite{zhou2020stabilizing}
\begin{align} \label{HinfBD}
\|F_i(s)\|_{\infty}^{[\omega_1,\omega_2]} &:= \underset{\omega \in [\omega_1,\omega_2]}{\sup}~|F_i(j\omega)|.
\end{align}
Observe that based on definitions \eqref{SCon} and \eqref{HinfBD}, the following inequalities:
\begin{align*}
    \|F_i(s)\|_{\infty}^{[\omega_1,\omega_2]} \le \|F_i(s)\|_{\infty} \le 1,
\end{align*}
are satisfied if \eqref{SCon} holds.

Defining the $\mathcal{H}_{\infty}$ norm over the predominant acceleration frequency boundaries of human-driven vehicles, we are now ready to state the main problem to be tackled in this work.

\begin{mypbm} \label{Pbm1}
    Given the CAV system in \eqref{CAVSys} governed by the control strategy in \eqref{DLC}, the predominant acceleration frequency boundaries of human-driven vehicles $\omega_2 > \omega_1 > 0$, and the following \textit{box constraints} (i.e., the lower and upper bounds) on the feedback and feedforward gains $\begin{bmatrix}
    k_1 & k_2 & k_3 & k_4
\end{bmatrix}$:
    \begin{subequations} \label{BC}
    \begin{empheq}[box=\widefbox]{align}
        & k_1^l \le k_1 \le k_1^u, \label{BC1}\\
        & k_2^l \le k_2 \le k_2^u,\label{BC2}\\
        & k_3^l \le k_3 \le k_3^u,\label{BC3}\\
        & k_4^l \le k_4 \le k_4^u,\label{BC4}
    \end{empheq}
    \end{subequations}
    synthesize an $\mathcal{H}_{\infty}^{[\omega_1,\omega_2]}$ optimal control strategy with an optimal $\mathcal{H}_{\infty}^{[\omega_1,\omega_2]}$ norm of $\gamma$ ($\gamma \le 1$). 
\end{mypbm}

Solving Problem \ref{Pbm1} enables us to effectively attenuate the stop-and-go disturbance amplification throughout the mixed vehicular platoon over the predominant acceleration frequency boundaries of human-driven vehicles. Furthermore, such a controller synthesis \textit{(i)} facilitates the effective stop-and-go disturbance attenuation for the scenario with a large communication delay, and \textit{(ii)} systematically incorporates the lower and upper bounds arising from the physics of the problem. It is noteworthy that these objectives are not addressable via the controller synthesis proposed by \cite{zhou2020stabilizing}.

\section{Main Results} \label{MRs}

To effectively investigate Problem \ref{Pbm1}, the current section is divided into three main parts: \textit{(a)} parameterized locally stabilizing feedback and feedforward gains subject to box constraints, \textit{(b)} string stability via the Padé approximation, and \textit{(c)} a near-optimal solution to Problem \ref{Pbm1}. By parameterizing locally stabilizing feedback and feedforward gains subject to box constraints, first, we systematically incorporate the local stability and the box constraints. Second, we additionally ensure the string stability via the Padé approximation. Finally, taking advantage of the previous two stages, we efficiently solve for a near-optimal solution to Problem \ref{Pbm1}.
\\To enhance the readability/tractability, we move all the proofs to Appendices \ref{App1}--\ref{App3}. 

\subsection{Parameterized locally stabilizing gains}

In this section, we elaborate on how to effectively guarantee local stability subject to the satisfaction of the box constraints via handy parameterizations. 

The set of feedback gains $\begin{bmatrix}
    k_1 & k_2 & k_3
\end{bmatrix}$ satisfying the local stability \eqref{LSC} can be parameterized via the parameters $\begin{bmatrix}
    x & y & z
\end{bmatrix}$ as
\begin{subequations} \label{PZ}
\begin{align}
    k_1(x) &= x,\\
    k_2(x,y) &= -\tau x + y,\\
    k_3(x,y,z) &= \frac{-Tx+y}{Ky}-z, 
\end{align}    
\end{subequations}
where $x$, $y$, and $z$ are all positive parameters.\\ Imposing box constraints \eqref{BC1}--\eqref{BC3} to the parameterization \eqref{PZ}, we obtain the following parameterization of the parameters $x$, $y$, and $z$ in \eqref{PZ}:
\begin{myprs} \label{Propo1}
The parameters $\begin{bmatrix}
    x & y & z
\end{bmatrix}$ in \eqref{PZ} satisfying box constraints \eqref{BC1}--\eqref{BC3} can be parameterized via the parameters $\begin{bmatrix}
    \psi_1 & \psi_2 & \psi_3
\end{bmatrix}$ as
\begin{subequations} \label{xyz}
    \begin{align}
        &x = \texttt{x}(\psi_1) = (1-\psi_1) x^l + \psi_1 x^u,\\
        &y = \texttt{y}(\psi_1,\psi_2) = (1-\psi_2) y^l_{\psi_1} + \psi_2 y^u_{\psi_1},\\
        &z = \texttt{z}(\psi_1,\psi_2,\psi_3) = (1-\psi_3) z^l_{\psi_1,\psi_2} + \psi_3 z^u_{\psi_1,\psi_2},\\
        & \textrm{with} \notag\\
        &x^l = \max \{\epsilon,k_1^l\},\\
    &x^u = k_1^u,\\
    &y^l_{\psi_1} = \max \bigg \{\epsilon, \tau \texttt{x}(\psi_1)+k_2^l,\frac{T \texttt{x}(\psi_1)}{-Kk_3^l+1-K\epsilon} \bigg \},\\
    &y^u_{\psi_1} = \tau \texttt{x}(\psi_1)+k_2^u,\\
    &z^l_{\psi_1,\psi_2} = \max \bigg \{\epsilon, \frac{-T \texttt{x}(\psi_1) + \texttt{y}(\psi_1,\psi_2)}{K \texttt{y}(\psi_1,\psi_2)}-k_3^u \bigg \},\\
    &z^u_{\psi_1,\psi_2} = \frac{-T \texttt{x}(\psi_1) + \texttt{y}(\psi_1,\psi_2)}{K \texttt{y}(\psi_1,\psi_2)} - k_3^l,
    \end{align}
\end{subequations}where $\psi_i \in [0,1]$ holds for all $i \in \{1,2,3\}$ and $\epsilon > 0$ is an infinitesimal value.
\end{myprs}
For the proof, the reader is referred to Appendix \ref{App1}. 

Similar to Proposition \ref{Propo1}, the feedforward gain $k_4$ satisfying box constraint \eqref{BC4} can simply be parameterized via the parameter $\psi_4$ as
\begin{align} \label{k4}
    k_4 &= (1-\psi_4)k_4^l + \psi_4 k_4^u, 
\end{align}
where $\psi_4 \in [0,1]$ holds. It is noteworthy that in parameterizations \eqref{xyz} and \eqref{k4}, the form of $\psi_i$'s can be chosen via any arbitrary sigmoid function, e.g., the logistic function
\begin{align*}
    \psi(\beta) &= \frac{1}{1+e^{-\zeta \beta}} \, ,  
\end{align*} 
where $\zeta > 0$ denotes the logistic growth rate. Merging \eqref{PZ}--\eqref{k4} along with sigmoid functions, we obtain the following parameterization of the locally stabilizing feedback and feedforward gains $\begin{bmatrix}
    k_1 & k_2 & k_3 & k_4
\end{bmatrix}$ subject to box constraints \eqref{BC}:
\begin{mycor} \label{Coro1}
    The locally stabilizing feedback and feedforward gains $\begin{bmatrix}
    k_1 & k_2 & k_3 & k_4
\end{bmatrix}$ subject to box constraints \eqref{BC} can be parameterized via the parameters $\begin{bmatrix}
    \kappa_1 & \kappa_2 & \kappa_3 & \kappa_4
\end{bmatrix}$ as
\begin{subequations} \label{FPara}
\begin{align}
    & k_1(\kappa_1) = \texttt{x}(\psi(\kappa_1)),\\
    & k_2(\kappa_1,\kappa_2) = -\tau \texttt{x}(\psi(\kappa_1)) + \texttt{y}(\psi(\kappa_1),\psi(\kappa_2)),\\
    & k_3(\kappa_1,\kappa_2,\kappa_3) = \frac{-T\texttt{x}(\psi(\kappa_1)) + \texttt{y}(\psi(\kappa_1),\psi(\kappa_2))}{K\texttt{y}(\psi(\kappa_1),\psi(\kappa_2))} \notag\\
    & - \texttt{z}(\psi(\kappa_1),\psi(\kappa_2),\psi(\kappa_3)),\\
    & k_4(\kappa_4) = (1-\psi(\kappa_4))k_4^l + \psi(\kappa_4) k_4^u,
\end{align}    
\end{subequations}where $\kappa_i \in \mathbb{R}$ holds for all $i \in \{1,2,3,4\}$ and $\texttt{x}()$, $\texttt{y}()$, and $\texttt{z}()$ represent the same functions expressed in \eqref{xyz}.
\end{mycor}

\begin{mylem} \label{Lem1}
    If the sufficient condition on the string stability \eqref{SCon} holds for the transfer function $F_i(s)$ in \eqref{TF}, then inequality \eqref{SSC3} holds for the control parameters $\begin{bmatrix}
    k_1 & k_2 & k_3 & k_4
\end{bmatrix}$.
\end{mylem}
The proof is in Appendix \ref{App2}. 

It can easily be verified that utilizing the necessary condition (associated with the string stability) stated by Lemma \ref{Lem1}, i.e., inequality \eqref{SSC3}, and noting that $K >0$ and $k_1 > 0$ hold, the feedforward gain $k_4$ can be parameterized via the parameters $\begin{bmatrix}
    x & y & z & w
\end{bmatrix}$ as
\begin{align} \label{k4p}
    k_4(x,y,z,w) &= \frac{\tau^2 x}{2} - \tau y + \frac{Tx}{Ky} + z + w,
\end{align}
where $x$, $y$, and $z$ are as expressed in \eqref{PZ} and $w$ is a non-negative parameter. Imposing box constraints \eqref{BC} to the parameterizations \eqref{PZ} and \eqref{k4p}, we obtain the following parameterization of the parameters $x$, $y$, $z$, and $w$ in \eqref{PZ} and \eqref{k4p}:
\begin{myprs} \label{Prop2}
    The parameters $\begin{bmatrix}
    x & y & z & w
\end{bmatrix}$ in \eqref{PZ} and \eqref{k4p} satisfying box constraints \eqref{BC} can be parameterized via the parameters $\begin{bmatrix}
    \psi_1 & \psi_2 & \psi_3 & \psi_4
\end{bmatrix}$ as
\begin{subequations} \label{wpar}
\begin{align}
    & ~x = \texttt{x}(\psi_1) = (1-\psi_1) x^l + \psi_1 x^u,\\
    & ~y = \texttt{y}(\psi_1,\psi_2) = (1-\psi_2) y^l_{\psi_1} + \psi_2 y^u_{\psi_1},\\
    & ~z = \texttt{z}(\psi_1,\psi_2,\psi_3) = (1-\psi_3) z^l_{\psi_1,\psi_2} + \psi_3 z^u_{\psi_1,\psi_2},\\
    & ~w = \texttt{w}(\psi_1,\psi_2,\psi_3,\psi_4) \notag \\ &= (1-\psi_4) w_{\psi_1,\psi_2,\psi_3}^l + \psi_4 w_{\psi_1,\psi_2,\psi_3}^u,\\
    & ~ \textrm{with} \notag\\
    & ~x^l = \max \{\epsilon,k_1^l\},\\
    & ~x^u = k_1^u,\\
    & ~y^l_{\psi_1} = \max \bigg \{\epsilon, \tau \texttt{x}(\psi_1)+k_2^l,\frac{T \texttt{x}(\psi_1)}{-Kk_3^l+1-K\epsilon},\notag \\ 
    & \frac{\xi(\texttt{x}(\psi_1)) + \sqrt{\xi(\texttt{x}(\psi_1))^2+\frac{4T \tau \texttt{x}(\psi_1)}{K}}}{2 \tau}  \bigg \},\\
    & ~y^u_{\psi_1} = \tau \texttt{x}(\psi_1)+k_2^u,\\
    & ~z^l_{\psi_1,\psi_2} = \max \bigg \{\epsilon, \frac{-T \texttt{x}(\psi_1) + \texttt{y}(\psi_1,\psi_2)}{K \texttt{y}(\psi_1,\psi_2)}-k_3^u \bigg \},\\
    & ~z^u_{\psi_1,\psi_2} = \min \bigg \{ \frac{-T \texttt{x}(\psi_1) + \texttt{y}(\psi_1,\psi_2)}{K \texttt{y}(\psi_1,\psi_2)} - k_3^l,\notag\\
    & \frac{-\tau^2 \texttt{x}(\psi_1)}{2} + \tau \texttt{y}(\psi_1,\psi_2) + \frac{-T \texttt{x}(\psi_1)}{K \texttt{y}(\psi_1,\psi_2)} + k_4^u \bigg \}, \\
    & ~w_{\psi_1,\psi_2,\psi_3}^l = \max \bigg \{0, \frac{-\tau^2 \texttt{x}(\psi_1)}{2} + \tau \texttt{y}(\psi_1,\psi_2) \notag\\
    & + \frac{-T \texttt{x}(\psi_1)}{K \texttt{y}(\psi_1,\psi_2)} - \texttt{z}(\psi_1,\psi_2,\psi_3) + k_4^l \bigg \},\\
    & ~w_{\psi_1,\psi_2,\psi_3}^u = \frac{-\tau^2 \texttt{x}(\psi_1)}{2} + \tau \texttt{y}(\psi_1,\psi_2) \notag \\
    & + \frac{-T \texttt{x}(\psi_1)}{K \texttt{y}(\psi_1,\psi_2)} - \texttt{z}(\psi_1,\psi_2,\psi_3) + k_4^u,
\end{align}    
\end{subequations}
where $\psi_i \in [0,1]$ holds for all $i \in \{1,2,3,4\}$ and $\xi(\texttt{x}(\psi_1)) = \frac{\tau^2 \texttt{x}(\psi_1)}{2} + \epsilon - k_4^u$.
\end{myprs}
See Appendix \ref{App3} for the proof.

Merging \eqref{PZ}, \eqref{k4p}, and \eqref{wpar} along with sigmoid functions, we obtain the following parameterization of the locally stabilizing feedback and feedforward gains $\begin{bmatrix}
    k_1 & k_2 & k_3 & k_4
\end{bmatrix}$ subject to box constraints \eqref{BC}, given in the following corrollary. 

\begin{mycor} \label{Coro2}
    The locally stabilizing feedback and feedforward gains $\begin{bmatrix}
    k_1 & k_2 & k_3 & k_4
\end{bmatrix}$ subject to box constraints \eqref{BC} can be parameterized via the parameters $\begin{bmatrix}
    \kappa_1 & \kappa_2 & \kappa_3 & \kappa_4
\end{bmatrix}$ as
\begin{subequations} \label{FPara2}
\begin{align}
    & k_1(\kappa_1) = \texttt{x}(\psi(\kappa_1)),\\
    & k_2(\kappa_1,\kappa_2) = -\tau \texttt{x}(\psi(\kappa_1)) + \texttt{y}(\psi(\kappa_1),\psi(\kappa_2)),\\
    & k_3(\kappa_1,\kappa_2,\kappa_3) = \frac{-T\texttt{x}(\psi(\kappa_1)) + \texttt{y}(\psi(\kappa_1),\psi(\kappa_2))}{K\texttt{y}(\psi(\kappa_1),\psi(\kappa_2))} \notag\\
    & - \texttt{z}(\psi(\kappa_1),\psi(\kappa_2),\psi(\kappa_3)),\\
    & k_4(\kappa_1,\kappa_2,\kappa_3,\kappa_4) = \frac{\tau^2 \texttt{x}(\psi(\kappa_1))}{2} - \tau \texttt{y}(\psi(\kappa_1),\psi(\kappa_2)) \notag \\
        & + \frac{T \texttt{x}(\psi(\kappa_1))}{K \texttt{y}(\psi(\kappa_1),\psi(\kappa_2))} + \texttt{z}(\psi(\kappa_1),\psi(\kappa_2),\psi(\kappa_3)) \notag \\
        & + \texttt{w}(\psi(\kappa_1),\psi(\kappa_2),\psi(\kappa_3),\psi(\kappa_4)),
\end{align}    
\end{subequations}where $\kappa_i \in \mathbb{R}$ holds for all $i \in \{1,2,3,4\}$ and $\texttt{x}()$, $\texttt{y}()$, $\texttt{z}()$, and $\texttt{w}()$ represent the same functions expressed in \eqref{wpar}.

\end{mycor}

\begin{mycor} \label{newcor}
Given the locally stabilizing feedback and feedforward gains $\begin{bmatrix}
    k_1 & k_2 & k_3 & k_4
\end{bmatrix}$ subject to box constraints \eqref{BC} and utilizing the parameterization \eqref{FPara2}, the corresponding parameters $\begin{bmatrix}
    \kappa_1 & \kappa_2 & \kappa_3 & \kappa_4
\end{bmatrix}$ can be extracted as
\begin{subequations} \label{kappar}
\begin{align}
    & \kappa_1(k_1) = \frac{1}{\zeta} \ln \bigg (\frac{\phi_1(k_1) - x^l}{x^u - \phi_1(k_1)} \bigg ),\\
    & \kappa_2(k_1,k_2,\kappa_1) = \frac{1}{\zeta} \ln \bigg (\frac{\phi_2(k_1,k_2) - y^l_{\psi(\kappa_1)}}{y^u_{\psi(\kappa_1)} - \phi_2(k_1,k_2)} \bigg ),\\
    & \kappa_3(k_1,k_2,k_3,\kappa_1,\kappa_2) = \notag\\ &\frac{1}{\zeta} \ln \bigg (\frac{\phi_3(k_1,k_2,k_3) - z^l_{\psi(\kappa_1),\psi(\kappa_2)}}{z^u_{\psi(\kappa_1),\psi(\kappa_2)} - \phi_3(k_1,k_2,k_3)} \bigg ),\\
    & \kappa_4(k_1,k_2,k_3,k_4,\kappa_1,\kappa_2,\kappa_3) = \notag \\
    & \frac{1}{\zeta} \ln \bigg (\frac{\phi_4(k_1,k_2,k_3,k_4) - w^l_{\psi(\kappa_1),\psi(\kappa_2),\psi(\kappa_3)}}{w^u_{\psi(\kappa_1),\psi(\kappa_2),\psi(\kappa_3)} - \phi_4(k_1,k_2,k_3,k_4)} \bigg ),\\
    & \textrm{with} \notag\\
    & \phi_1(k_1) = k_1,\\
    & \phi_2(k_1,k_2) = \tau k_1 + k_2,\\
    & \phi_3(k_1,k_2,k_3) = \frac{-T k_1}{K(\tau k_1 + k_2)} + \frac{1}{K} - k_3,\\
    & \phi_4(k_1,k_2,k_3,k_4) = \frac{\tau^2}{2}k_1 + \tau k_2 + k_3 + k_4 - \frac{1}{K},
\end{align}
\end{subequations}where 
\begin{align*}
    & y^l_{\psi(\kappa_1)},~y^u_{\psi(\kappa_1)},~z^l_{\psi(\kappa_1),\psi(\kappa_2)},~z^u_{\psi(\kappa_1),\psi(\kappa_2)},\\
    & w^l_{\psi(\kappa_1),\psi(\kappa_2),\psi(\kappa_3)},~w^u_{\psi(\kappa_1),\psi(\kappa_2),\psi(\kappa_3)},
\end{align*}
are evaluated by replacing $\psi_i$ (in \eqref{wpar}) with $\psi(\kappa_i)$ for all $i \in \{1,2,3\}$.
\end{mycor}

As a summary, \textit{(i)} Corollary \ref{Coro2} will be utilized as a cornerstone to parameterize the locally stabilizing feedback and feedforward gains $\begin{bmatrix}
    k_1 & k_2 & k_3 & k_4
\end{bmatrix}$ via the parameters $\begin{bmatrix}
    \kappa_1 & \kappa_2 & \kappa_3 & \kappa_4
\end{bmatrix}$, and \textit{(ii)} Corollary \ref{newcor} will facilitate the extraction of the parameters $\begin{bmatrix}
    \kappa_1 & \kappa_2 & \kappa_3 & \kappa_4
\end{bmatrix}$ given the locally stabilizing feedback and feedforward gains $\begin{bmatrix}
    k_1 & k_2 & k_3 & k_4
\end{bmatrix}$. The former is a useful feature in the case of searching for the near-optimal solution to Problem \ref{Pbm1} while the latter is a key tool to the case of opting (extracting) an initial feasible point for the main optimization problem associated with Problem \ref{Pbm1}. 

\subsection{String stability via the Padé approximation}

In this section, we shed light on incorporating the string stability into the locally stabilizing box-constrained gains parameterized in the previous section. 

As mentioned earlier, one needs to overcome the complexity of the communication delay in the $\mathcal{H}_{\infty}$ controller synthesis. In that regard, unlike the Taylor series approximation utilized by \cite{zhou2020stabilizing}, we employ the Padé approximation \cite{golub1989matrix} approach to approximate the transfer function $F_i(s)$ in \eqref{TF}, namely, $\hat{F}_i(s)$. To that end, we replace the exponential term associated with the communication delay $\theta$ in $F_i(s)$ in \eqref{TF}, i.e., $e^{-\theta s}$ with its Padé approximation which poses the following rational form:
\begin{align} \label{PAPP}
    e^{-\theta s} & \approx \frac{P^{Pad\acute{e}}(s)}{Q^{Pad\acute{e}}(s)},
\end{align}
where $P^{Pad\acute{e}}(s)$ and $Q^{Pad\acute{e}}(s)$ denote the $N$-th order ($N$: Padé approximation order) numerator and denominator polynomials, respectively. Inspired by the sufficient condition \eqref{SCon}, to more accurately ensure the string stability of the CAV system in \eqref{CAVSys} governed by the control strategy in \eqref{DLC} in the presence of communication delay, we utilize the following condition:
\begin{align} \label{ASC}
\|\hat{F}_i(s)\|_{\infty} \le 1.
\end{align}
Notice that in our case, deriving a set of inequalities similar to the inequalities \eqref{SSC} proposed by \href{https://www.sciencedirect.com/science/article/pii/S0191261518311482?casa_token=NzZJbHpjaSwAAAAA:UPUagh0REEv0nvScPTw_PEmOfgaPnfwXx1TWtitr4Hg7rL2FlOfbiWUy3DdbkHuXWjxaGkC2}{Proposition $2$} in \cite{zhou2020stabilizing} is impossible as we do not limit ourselves to the case that $\theta$ is sufficiently small in the current paper. Specifically, the authors in \cite{zhou2020stabilizing} take advantage of the specific properties of the corresponding quartic polynomial $p\omega^4 + q\omega^2 + r \ge 0$ resulting from the aforementioned simplifying assumption regarding the $\theta$ that is no longer applicable to our case.

Centering around \eqref{ASC}, we incorporate the string stability into the locally stabilizing box-constrained gains by utilizing the Padé approximation \eqref{PAPP}. Also, later on, we empirically observe that the Padé approximation provides a more accurate approximation than the Taylor series approximation, leading to a potentially better near-optimal solution. We emphasize that although there exists an analytical explicit formula for $F_i(j\omega)$ in \eqref{SCon} and \eqref{HinfBD} (according to \eqref{Fabs}), utilizing the approximate forms of \eqref{SCon} and \eqref{HinfBD}, i.e., \eqref{PAPP}, is inevitable as \eqref{SCon} and \eqref{HinfBD} cannot be utilized directly due to the computational complexity of computing $\|F_i(s)\|_{\infty}$ and $\|F_i(s)\|_{\infty}^{[\omega_1,\omega_2]}$, respectively. Then, we utilize the Padé approximation-based counterparts $\|\hat{F}_i(s)\|_{\infty}$ and $\|\hat{F}_i(s)\|_{\infty}^{[\omega_1,\omega_2]}$ in the next section to effectively solve Problem \ref{Pbm1} for a near-optimal solution.

\subsection{A near-optimal solution to Problem \ref{Pbm1}}

Here, we propose a two-stage procedure to solve Problem \ref{Pbm1} for a near-optimal solution. Such two stages are as follows:
\begin{enumerate}
    \item Finding an initial stabilizing feasible point satisfying box constraints \eqref{BC} and both local stability \eqref{LSC} and string stability \eqref{SCon}.
    \item Solving Problem \ref{Pbm1} for a near-optimal solution starting from the obtained initial stabilizing feasible point in the first stage. 
\end{enumerate}

Substituting the parameterized locally stabilizing feedback and feedforward gains \eqref{FPara2} provided by Corollary \ref{Coro2} into $F_i(s)$ in \eqref{TF}, we obtain the parameterized $F(s;\kappa)$ and denote its Padé approximation by $\hat{F}(s;\kappa)$. Then, defining the optimization variable $\kappa$ as $\kappa := \begin{bmatrix}
    \kappa_1 & \kappa_2 & \kappa_3 & \kappa_4
\end{bmatrix}^\top$, we consider the following optimization problem:
\begin{subequations} \label{OPPBM}
\begin{align}
    \underset{\kappa \in \mathbb{R}^4}{\mathrm{Minimize}}~&\|\hat{F}(s;\kappa)\|_{\infty}^{[\omega_1,\omega_2]}, \label{OPPBM1}
    \\ \mathrm{subject~to:}~&\|\hat{F}(s;\kappa)\|_{\infty} \le 1, \label{OPPBM2}
\end{align}
\end{subequations}
to solve Problem \ref{Pbm1} for a near-optimal solution (Remember that $\|\hat{F}(s;\kappa)\|_{\infty}^{[\omega_1,\omega_2]}$ denotes the $\mathcal{H}_{\infty}^{[\omega_1,\omega_2]}$ norm of $\hat{F}(s;\kappa)$). The aforementioned two-stage procedure takes the following form:
\begin{enumerate}
    \item We first search for an initial stabilizing feasible point $\kappa^0$ that satisfies constraint \eqref{OPPBM2}.
    \item We then solve optimization problem \eqref{OPPBM} for a near-optimal solution $\kappa^{\ast}$ starting from the obtained initial stabilizing feasible point in the first stage, i.e., $\kappa^0$. Finally, one can compute $k^{\ast}$ by substituting $\kappa^{\ast}$ into the parameterization \eqref{FPara2} provided by Corollary \ref{Coro2} as $k^{\ast} = k(\kappa^{\ast})$.
\end{enumerate}In the sequel, we delve into each stage thoroughly. 

\subsubsection{First stage} Notably, one can take advantage of \eqref{kappar} provided by Corollary \ref{newcor} to extract $\kappa^0$ from $k^0$. Such a fact motivates us to: first, alternatively search for an initial stabilizing feasible point $k^0$ and second, extract the corresponding initial stabilizing feasible point $\kappa^0$ from $k^0$. To that end, first, we consider the following parameterization:
\begin{subequations} \label{KPR}
\begin{align}
    k_1(\mu_1) &= (1-\rho(\mu_1)) \max \{\epsilon ,k_1^l \} + \rho(\mu_1) k_1^u,\\
    k_2(\mu_2) &= (1-\rho(\mu_2))k_2^l + \rho(\mu_2) k_2^u,\\
    k_3(\mu_3) &= (1-\rho(\mu_3))k_3^l + \rho(\mu_3) k_3^u,\\
    k_4(\mu_4) &= (1-\rho(\mu_4))k_4^l + \rho(\mu_4) k_4^u,
\end{align}    
\end{subequations}
with
\begin{align} \label{rhofun}
    \rho(\beta) &= \frac{1}{1 + \nu \beta^2},
\end{align}
where $\nu > 0$ can arbitrarily be chosen and second, defining $\mu := \begin{bmatrix}
    \mu_1 & \mu_2 & \mu_3 & \mu_4 
\end{bmatrix}^\top$ and substituting the parameterized feedback and feedforward gains \eqref{KPR} into $F_i(s)$ in \eqref{TF}, we obtain the parameterized $G(s;\mu)$ and denote its Padé approximation by $\hat{G}(s;\mu)$.

Then, to search for an initial stabilizing feasible point $\mu^{0}$ that satisfies $\|\hat{G}(s;\mu)\|_{\infty} \le 1$ (an equivalent constraint to constraint \eqref{OPPBM2}), we solve the following standard $\mathcal{H}_{\infty}$ optimization problem:
\begin{align} \label{SDHIOP}
    \underset{\mu \in \mathbb{R}^4}{\mathrm{Minimize}}~&\|\hat{G}(s;\mu)\|_{\infty}, 
\end{align}
for $\mu^0$, utilizing a well-developed standard $\mathcal{H}_{\infty}$ problem solver built upon \cite{apkarian2006nonsmooth}, i.e., \texttt{hinfstruct} solver proposed by \cite{gahinet2011structured,gahinet2011decentralized}. Afterwards, according to the parameterization \eqref{KPR}, we get $k^0 = k(\mu^0)$ and plug $k^0$ to \eqref{kappar} provided by Corollary \ref{newcor} to extract $\kappa^0$. 

\begin{myrem}
    It is noteworthy that due to the fundamental limitations of the implementation of \texttt{hinfstruct}, one cannot utilize the most representative sophisticated parameterization \eqref{FPara2} provided by Corollary \ref{Coro2} at this stage, to solve \eqref{SDHIOP} for $\mu^0$. That is why we alternatively employ the simple parameterization \eqref{KPR} along with \eqref{rhofun} to that end.
\end{myrem}

\subsubsection{Second stage} Motivated by \eqref{ASC}, let us define the following function:
\begin{align} \label{Branch}
    h(\kappa) &:= \left\{
\begin{array}{ll}
       \|\hat{F}(s;\kappa)\|_{\infty}^{[\omega_1,\omega_2]} & \|\hat{F}(s;\kappa)\|_{\infty} \le 1 \\
      \alpha & \|\hat{F}(s;\kappa)\|_{\infty} > 1\\
\end{array}, 
\right. 
\end{align}
where $\alpha > 1$ can arbitrarily be chosen. Now, we can alternatively solve the following unconstrained optimization problem:
\begin{align} \label{UNCOPPBM}
    \underset{\kappa \in \mathbb{R}^4}{\mathrm{Minimize}}~&h(\kappa), 
\end{align}
to find a solution $\kappa^{\ast}$ to optimization problem \eqref{OPPBM}. Finally, one can compute $k^{\ast}$ by substituting $\kappa^{\ast}$ into the parameterization \eqref{FPara2} provided by Corollary \ref{Coro2} as $k^{\ast} = k(\kappa^{\ast})$. Regarding the initialization, we utilize $\kappa^0$ obtained from the first stage. Furthermore, due to the non-convex and non-smooth nature of the function $h(\tau)$ in \eqref{Branch}, we need to utilize non-convex and non-smooth optimization tools to solve the unconstrained optimization problem \eqref{UNCOPPBM}. For instance, we can employ \texttt{fminsearch} solver developed based on Nelder--Mead simplex method \cite{lagarias1998convergence} in that regard. It is noteworthy that given a non-convex non-smooth optimization problem, finding a globally optimal solution is generally a challenging task. As a result, we utilize the term \textit{near-optimal} instead of the term \textit{optimal} in the current paper.

Procedure \ref{proc:TSA} summarizes the two-stage $\mathcal{H}_{\infty}$ controller synthesis procedure.

\setlength{\floatsep}{5pt}{
\begin{algorithm}[!ht]
\caption{\textit{Two-stage $\mathcal{H}_{\infty}$ Controller Synthesis}}\label{proc:TSA}
\DontPrintSemicolon
\textbf{Input:} $\tau$,$T$,$K$,$\theta$,$ \omega_1$,$\omega_2$,$\{k_i^l\}_{i=1}^{4}$,$\{k_i^u\}_{i=1}^{4}$,$\alpha$,$\zeta$,$\nu$,$N$

\hspace{0.05in} \textit{First stage}:

\hspace{0.1in} Construct $G(s;\mu)$ via the parameterization \eqref{KPR}.

\hspace{0.1in} Get $\hat{G}(s;\mu)$ via the Padé approximation of $G(s;\mu)$.

\hspace{0.1in} Solve \eqref{SDHIOP} for $\mu^0$ via \texttt{hinfstruct} solver.

\hspace{0.1in} Compute $k^0 = k(\mu^0)$ via the parameterization \eqref{KPR}.

\hspace{0.1in} Extract $\kappa^0$ from $k^0$ via \eqref{kappar}.

\hspace{0.05in} \textit{Second stage}:

\hspace{0.1in} Construct $F(s;\kappa)$ via the parameterization \eqref{FPara2}.

\hspace{0.1in} Get $\hat{F}(s;\kappa)$ via the Padé approximation of $F(s;\kappa)$.

\hspace{0.1in} Initialize \eqref{UNCOPPBM} with $\kappa^0$ obtained from the first stage.

\hspace{0.1in} Solve \eqref{UNCOPPBM} for $\kappa^{\ast}$ via \texttt{fminsearch} solver.

\hspace{0.1in} Compute $k^{\ast} = k(\kappa^{\ast})$ via the parameterization \eqref{FPara2}. 

\textbf{Output:} $k^{\ast}$.
\end{algorithm}
}

\begin{myrem}
    Observe that for the parameterized $F(s;\kappa)$, on the one hand, according to the supremum-based definition of the $\mathcal{H}_{\infty}$ norm expressed in \eqref{SCon}, we have $\|F(s;\kappa)\|_{\infty} \ge 1$ based on $\lim_{\omega \to 0^+} F(j\omega;\kappa) = 1$, $\lim_{\omega \to 0^+} F'(j\omega;\kappa) = 0$, and $\lim_{\omega \to 0^+} F''(j\omega;\kappa) \le 0$ (See Appendix \ref{App2} for more details). On the other hand, according to \eqref{SCon}, we impose $\|F(s;\kappa)\|_{\infty} \le 1$ to obtain the parameterized locally stabilizing feedback and feedforward gains that guarantee the string stability. Thus, we conclude that for such feedback and feedforward gains $\|F(s;\kappa)\|_{\infty} = 1$ holds. Particularly, for $\kappa^0$ and $\kappa^{\ast}$, the equalities $\|F(s;\kappa)\|_{\infty} = 1$ and $\|F(s;\kappa)\|_{\infty} = 1$ are satisfied, respectively.   
\end{myrem}

\begin{myrem}
    The first stage presented in Procedure \ref{proc:TSA} can be devised differently. In other words, various alternatives exist to choose an appropriate $\kappa^0$. As an effective alternative, one can think of sampling the $4$-dimensional search space $\mathbb{R}^4$ equipped with a pre-assumed distribution (e.g., uniform distribution). Denoting the set of samples $\tilde{\kappa}$ by $\mathcal{S}$, for any sample $\tilde{\kappa} \in \mathcal{S}$, we can check that if $\|F(s;\tilde{\kappa})\|_{\infty} \le 1$ holds or not. Collecting all those samples satisfying $\|F(s;\tilde{\kappa})\|_{\infty} \le 1$ and sorting the corresponding $\mathcal{H}_{\infty}^{[\omega_1,\omega_2]}$ norm values, i.e., $h(\tilde{\kappa}) = \|\hat{F}(s;\tilde{\kappa})\|_{\infty}^{[\omega_1,\omega_2]}$, in an ascending order, we can set $\kappa^0$ as $\kappa^0 = \argmin_{\tilde{\kappa} \in \mathcal{S}} h(\tilde{\kappa})$. Notice that at the expense of a denser sampling (higher computational time), one expects to obtain a more representative choice of such $\kappa^0$ leading to a better near-optimal solution $k^{\ast}$. However, the first stage presented in Procedure \ref{proc:TSA} significantly saves computational time while mostly generating a high-quality choice of $\kappa^0$ leading to an acceptable near-optimal solution $k^{\ast}$. 
\end{myrem}

\section{Numerical Simulations} \label{secNS}

Throughout this section, we assess the effectiveness of the theoretical results presented in the paper. Depending on how small the communication delay is, the section is divided into two main parts: \textit{(a)} Case $1$: a sufficiently small communication delay $\theta$ and \textit{(b)} Case $2$: a large communication delay $\theta$.   

In Case $1$, since the communication delay $\theta$ is sufficiently small, we can set the unconstrained $\mathcal{H}_{\infty}$ controller synthesis proposed in \cite{zhou2020stabilizing}, namely, $k^{\mathrm{unc}}$, as a benchmark for comparative analysis purposes. In that regard, to have a fair comparison, we impose the box constraints on $k^{\ast}$, based on $\|k^{\mathrm{unc}}\|_{\ell_{\infty}}$. In other words, we consider a hypercube that encompasses $k^{\mathrm{unc}}$. In Case $2$, since $\theta$ is not sufficiently small anymore, the Taylor series approximation-based method proposed by \cite{zhou2020stabilizing} becomes unusable.

We conduct all the numerical experiments in MATLAB R$2022$b on a MacBook Pro with a $3.1$ GHz Intel Core i$5$ and memory $8$ GB $2133$ MHz. To create the continuous-time transfer function models, we employ \texttt{tf}. We also utilize \texttt{getPeakGain} (developed built upon \cite{bruinsma1990fast}) to evaluate the $\mathcal{H}_{\infty}^{[\omega_1,\omega_2]}$/$\mathcal{H}_{\infty}$ norm values. Regarding the Padé approximation, we utilize \texttt{pade} equipped with $N$ which denotes the Padé approximation order. To utilize \texttt{hinfstruct}, we take advantage of \texttt{realp}, creating real-valued tunable parameters $\mu_1$, $\mu_2$, $\mu_3$, and $\mu_4$. Furthermore, we initialize those real-valued parameters via \texttt{rand}.  

\begin{table}[t]
    \centering
    \caption{Value setting for the experimental $\mathcal{H}_{\infty}$ controller synthesis \cite{zhou2020stabilizing}. Time quantities are all in seconds.}
\label{tab:my_label}
    \begin{tabular}{|c|c|}
        \hline
        Parameter & Value \\
        \hline
        $\tau$ & $1$\\
        \hline
        $T$ & $0.45$\\
        \hline
        $K$ & $1$\\
        \hline
        $\theta$ & $0.1$\\
        \hline
        $\omega_2$ & $2.5$\\
        \hline
    \end{tabular}
\end{table}

\subsection{Case 1: a sufficiently small communication delay}
In this section, considering the case of sufficiently small $\theta$, we conduct a comparative analysis between the unconstrained and constrained $\mathcal{H}_{\infty}$ syntheses $k^{\mathrm{unc}}$ \cite{zhou2020stabilizing} and $k^{\ast}$.

\begin{figure}[!ht]
    \centering
    \includegraphics[scale = 0.22]{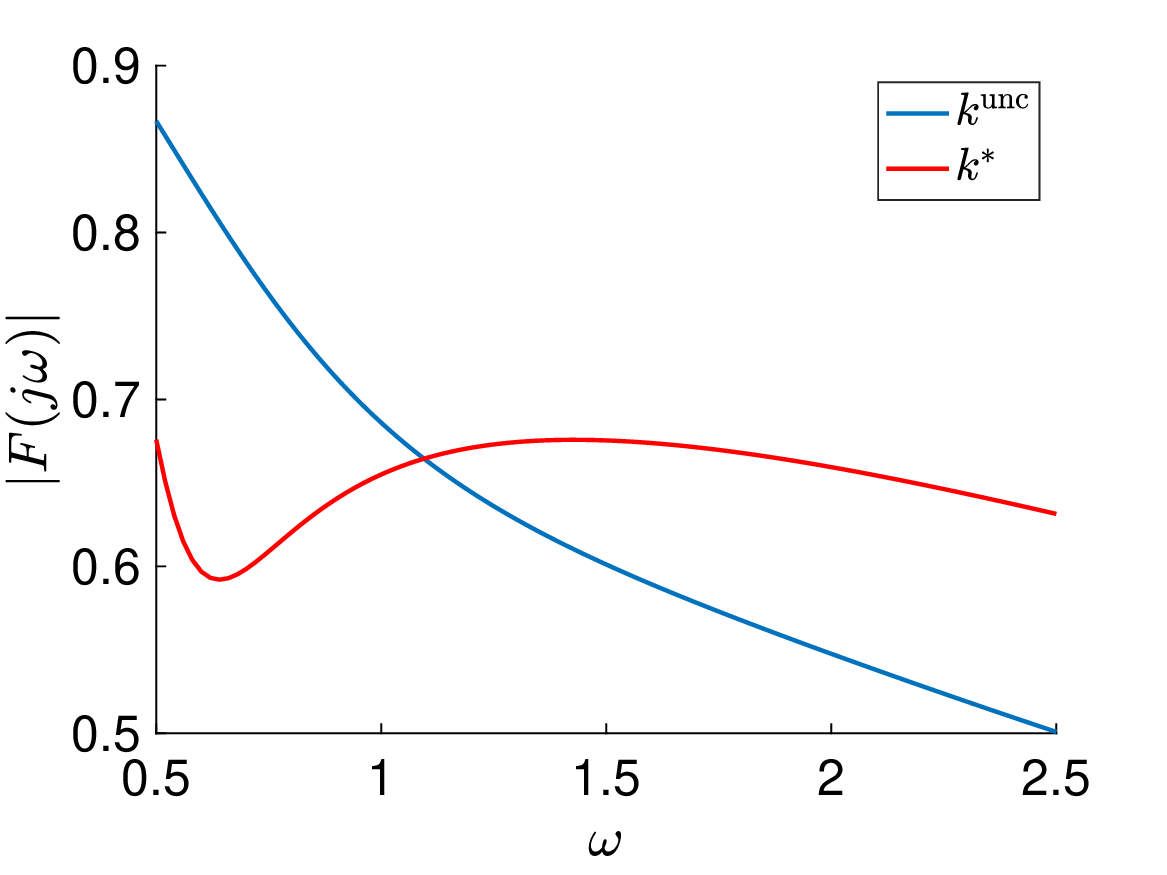}
    \includegraphics[scale = 0.22]{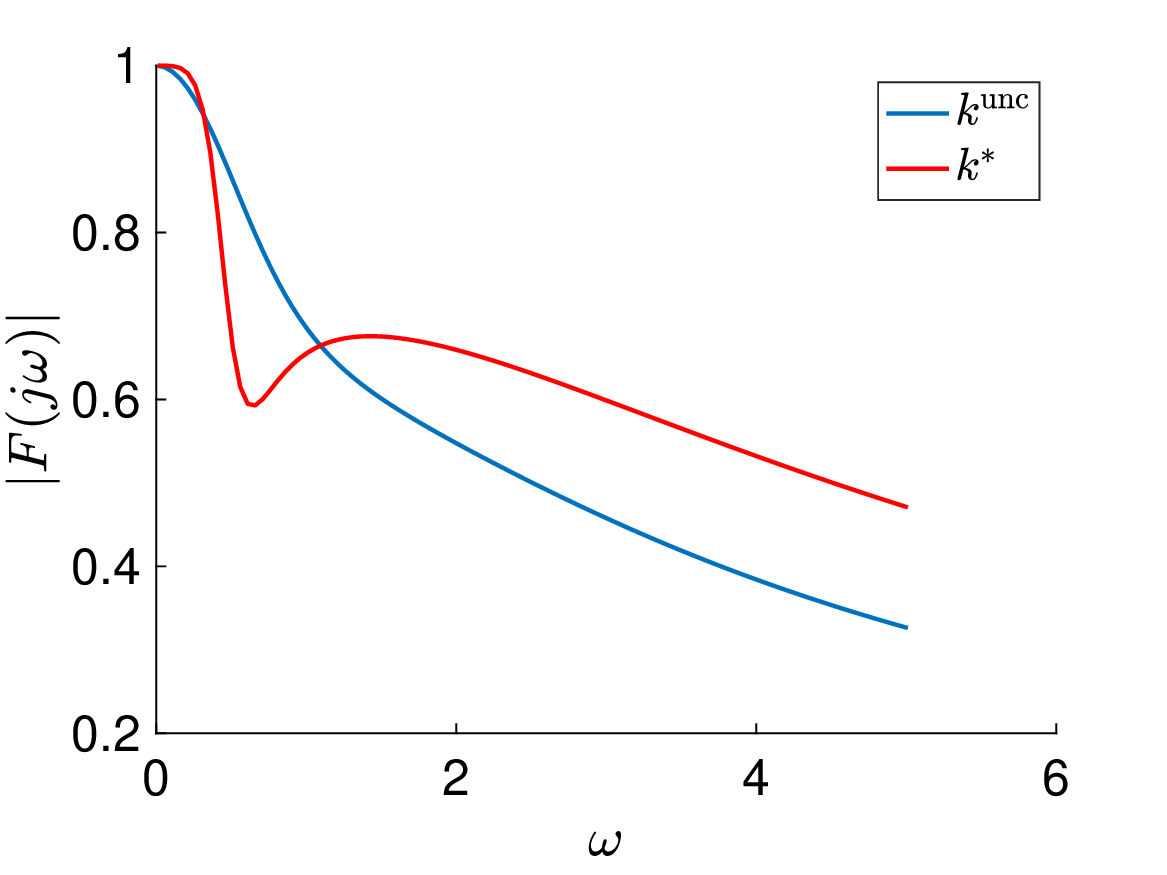}
    
    \caption{The $\omega$-dependent $|F(j\omega)|$ values corresponding to the unconstrained and constrained $\mathcal{H}_{\infty}$ syntheses $k^{\mathrm{unc}}$ and $k^{\ast}$ with $\theta = 0.1$ and $\omega_1 = 0.5$ for $\omega \in [\omega_1,\omega_2]$ on the left and $\omega \in [0.01,5.01]$ on the right.}
    \label{fig1}
\end{figure}

\begin{figure}[!ht]
    \centering

    \includegraphics[scale = 0.22]{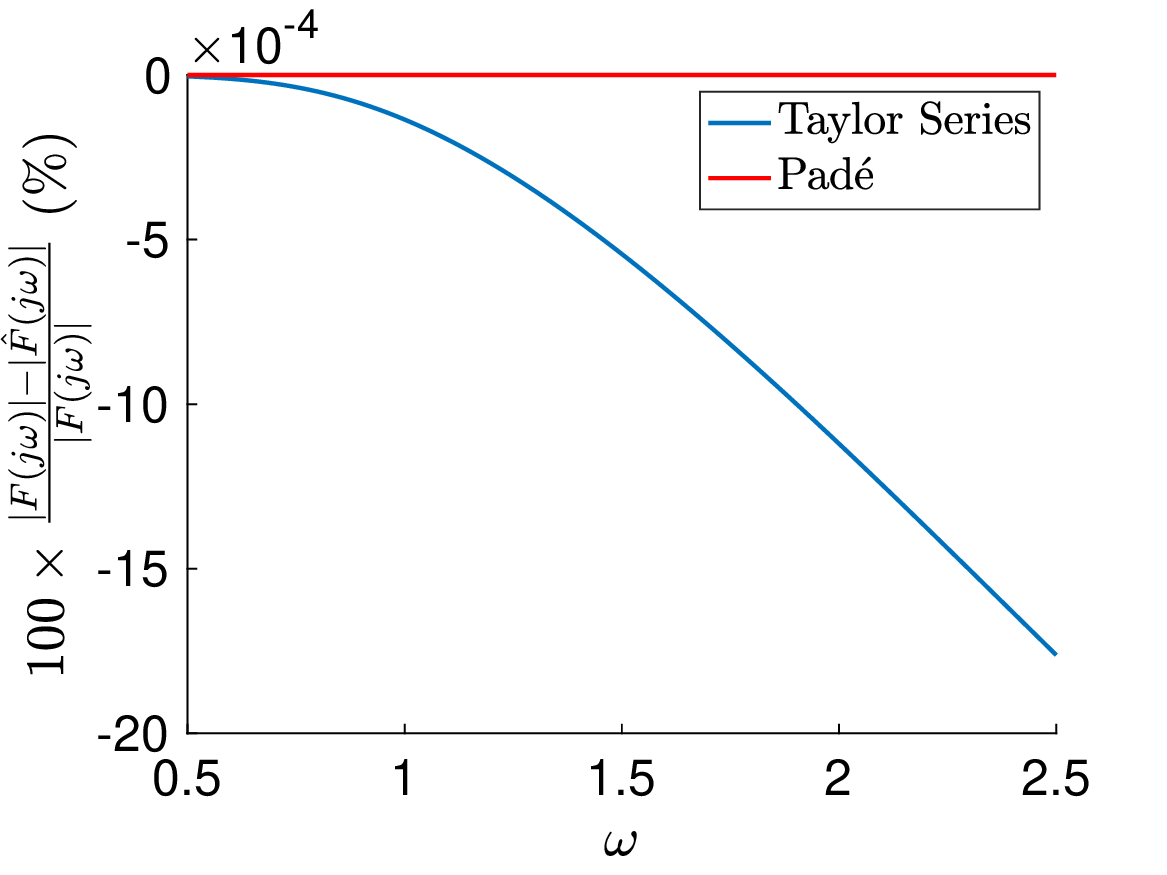}
    \includegraphics[scale = 0.22]{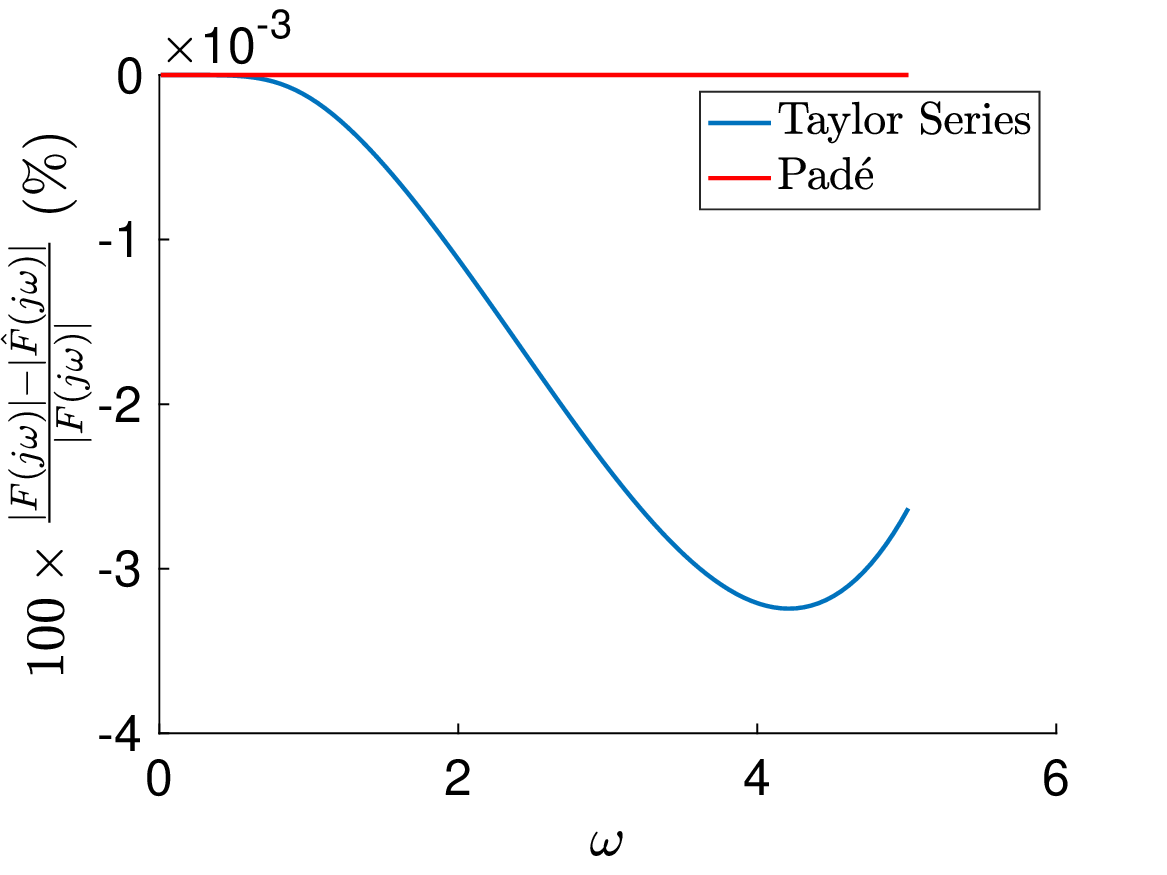}
    
    \caption{The relative error percentages $100 \times \frac{|F(j\omega)|-|\hat{F}(j\omega)|}{|F(j\omega)|}$ (\%) associated with the Taylor series approximation-based method and the Padé approximation-based method with $\theta = 0.1$ and $\omega_1 = 0.5$ for $\omega \in [\omega_1,\omega_2]$ on the left and $\omega \in [0.01,5.01]$ on the right.}
    \label{fig15}
\end{figure}

Similar to \cite{zhou2020stabilizing}, we consider the value setting in Tab. \ref{tab:my_label} for the experimental $\mathcal{H}_{\infty}$ controller synthesis. Time quantities are all in seconds. Setting $\omega_1 = 0.5$, $k^{l} = -\|k^{\mathrm{unc}}\|_{\ell_{\infty}} \begin{bmatrix}
    0 & 1 & 1 & 1
\end{bmatrix}^\top$, $k^{u} = \|k^{\mathrm{unc}}\|_{\ell_{\infty}} \mathbf{1}$, $\alpha = 1.05$, $\zeta = 5$, $\nu = 5$, $N = 5$ and running Procedure \ref{proc:TSA}, we obtain the two-stage $\mathcal{H}_{\infty}$ controller synthesis illustrated by Tab. \ref{tab:my_label2} for which the $\mathcal{H}_{\infty}^{[\omega_1,\omega_2]}$ norm values corresponding to the unconstrained and constrained $\mathcal{H}_{\infty}$ syntheses $k^{\ast}$ and $k^{\mathrm{unc}}$ \cite{zhou2020stabilizing} are $\|F(s)\|_{\infty}^{[\omega_1,\omega_2]} = 0.6758$ and $\|F(s)\|_{\infty}^{[\omega_1,\omega_2]} = 0.8667$, respectively. Accordingly, Fig. \ref{fig1} visualizes the $\omega$-dependent $|F(j\omega)|$ values corresponding to the unconstrained and constrained $\mathcal{H}_{\infty}$ syntheses $k^{\mathrm{unc}}$ and $k^{\ast}$ with $\theta = 0.1$ and $\omega_1 = 0.5$ for $\omega \in [\omega_1,\omega_2]$ on the left and $\omega \in [0.01,5.01]$ on the right. According to Tab. \ref{tab:my_label2}, $\lambda(A + Bk^{\ast}) = \{-8.0215,-0.4595 - 0.3606 j, -0.4595 + 0.3606 j\}$, and Fig. \ref{fig1}, we observe that box constraints \eqref{BC} and both local stability \eqref{LSC} and string stability \eqref{SCon} are satisfied for $k^{\ast}$. Also, Fig. \ref{fig15} depicts the relative error percentages $100 \times \frac{|F(j\omega)|-|\hat{F}(j\omega)|}{|F(j\omega)|}$ (\%) associated with the Taylor series approximation-based method and the Padé approximation-based method with $\theta = 0.1$ and $\omega_1 = 0.5$ for $\omega \in [\omega_1,\omega_2]$ on the left and $\omega \in [0.01,5.01]$ on the right. As Fig. \ref{fig15}, the Padé approximation-based method attains a more accurate approximation than the Taylor series approximation-based counterpart.

\begin{table}[t]
    \centering
    \caption{The two-stage $\mathcal{H}_{\infty}$ controller synthesis for the value setting in Tab. \ref{tab:my_label}, $\omega_1 = 0.5$, $k^{l} = -\|k^{\mathrm{unc}}\|_{\ell_{\infty}} \begin{bmatrix}
    0 & 1 & 1 & 1
\end{bmatrix}^\top$, $k^{u} = \|k^{\mathrm{unc}}\|_{\ell_{\infty}} \mathbf{1}$, $\alpha = 1.05$ , $\zeta = 5$, $\nu = 5$, $N = 5$. $k^{\mathrm{unc}}$: the unconstrained $\mathcal{H}_{\infty}$ controller synthesis \cite{zhou2020stabilizing}.}
\label{tab:my_label2}
    \begin{tabular}{|c|c|}
        \hline
        Vector & Value \\
        \hline
        $k^{\ast}$ & $\begin{bmatrix}
    0.4212 & 0.4775 & -1.0078 & 1.3197
\end{bmatrix}^\top$\\
        \hline
        $\kappa^{\ast}$ & $\begin{bmatrix}
    -0.1516 & -0.0237 & 1.7065 & -0.7647
\end{bmatrix}^\top$\\
        \hline
        $\kappa^0$ & $\begin{bmatrix}
    0.0918 & -0.0378 & -0.2983 & -0.1611
\end{bmatrix}^\top$\\
        \hline
        $k^0$ & $\begin{bmatrix}
    0.8089 & 0.3191 & 0.3611 & 0.3492
\end{bmatrix}^\top$\\
        \hline
        $\mu^0$ & $\begin{bmatrix}
    0.3555 & 0.3495 & 0.3377 & 0.3411
\end{bmatrix}^\top$\\
        \hline
         $k^{\mathrm{unc}}$ & $\begin{bmatrix}
    0.9200 & 1.3200 & -0.9200 & 0.7200
\end{bmatrix}^\top$\\
        \hline
    \end{tabular}
\end{table}

\begin{table}[t]
    \centering
    \caption{The $\omega_1$-dependency of the $\mathcal{H}_{\infty}^{[\omega_1,\omega_2]}$ norm values corresponding to the unconstrained and constrained $\mathcal{H}_{\infty}$ syntheses $k^{\mathrm{unc}}$ \cite{zhou2020stabilizing} and $k^{\ast}$ for predominant acceleration frequency boundary of human-driven vehicles $\omega_1 \in \{0.1,0.3,0.5,0.7\}$.}
\label{tab:my_label3}
    \begin{tabular}{|c|c|c|}
        \hline
        $\omega_1$ & $k$ & $\|F(s)\|_{\infty}^{[\omega_1,\omega_2]}$ \\
        \hline
        0.1 & $k^{\mathrm{unc}}$ & $0.9739$\\
        \hline
        0.1 & $k^{\ast}$ & \cellcolor{lightgray}$0.9628$\\
        \hline
        0.3 & $k^{\mathrm{unc}}$ & $0.9001$\\
        \hline
        0.3 & $k^{\ast}$ & \cellcolor{lightgray}$0.8207$\\
        \hline
        0.5 & $k^{\mathrm{unc}}$ & $0.8667$\\
        \hline
        0.5 & $k^{\ast}$ & \cellcolor{lightgray}$0.6758$\\
        \hline
        0.7 & $k^{\mathrm{unc}}$ & $0.7304$\\
        \hline
        0.7 & $k^{\ast}$ & \cellcolor{lightgray}$0.5669$\\
        \hline
    \end{tabular}
\end{table}

\begin{figure}[!ht]
    \centering
    \includegraphics[scale = 0.22]{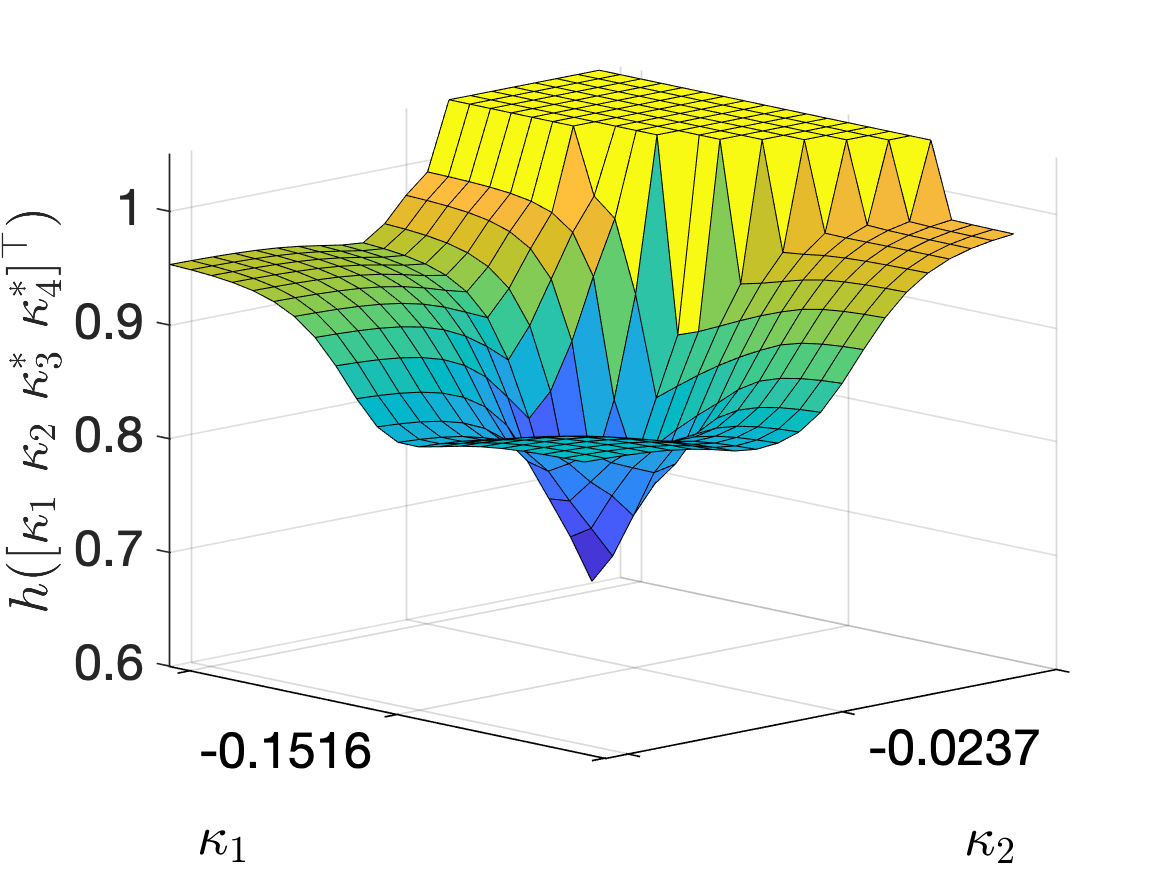}
    
    \caption{The $3$D search space (fixing $\kappa_3 = \kappa_3^{\ast}$ and $\kappa_4 = \kappa_4^{\ast}$) corresponding to the constrained $\mathcal{H}_{\infty}$ controller synthesis $k^{\ast}$ with $\theta = 0.1$ and $\omega_1 = 0.5$.}
    \label{fig2}
\end{figure}

Fig. \ref{fig2} visualizes the $3$D search space (fixing $\kappa_3 = \kappa_3^{\ast}$ and $\kappa_4 = \kappa_4^{\ast}$) corresponding to the constrained $\mathcal{H}_{\infty}$ controller synthesis $k^{\ast}$ with $\theta = 0.1$ and $\omega_1 = 0.5$.

For predominant acceleration frequency boundary of human-driven vehicles $\omega_1 \in \{0.1,0.3,0.5,0.7\}$, Tab. \ref{tab:my_label3} reflects the $\omega_1$-dependency of the $\mathcal{H}_{\infty}^{[\omega_1,\omega_2]}$ norm values corresponding to the unconstrained and constrained $\mathcal{H}_{\infty}$ syntheses $k^{\mathrm{unc}}$ and $k^{\ast}$. According to Tab. \ref{tab:my_label3}, the Padé approximation-based box-constrained solution outperforms the Taylor series approximation-based unconstrained solution \cite{zhou2020stabilizing} in terms of the $\mathcal{H}_{\infty}^{[\omega_1,\omega_2]}$ norm. As another observation that holds for both unconstrained and constrained $\mathcal{H}_{\infty}$ syntheses $k^{\mathrm{unc}}$ and $k^{\ast}$, the smaller predominant acceleration frequency boundary of human-driven vehicles $\omega_1$, the larger $\mathcal{H}_{\infty}^{[\omega_1,\omega_2]}$ norm. Such a monotonous trend is aligned with the fact that decreasing the value of predominant acceleration frequency boundary of human-driven vehicles $\omega_1$ in \eqref{HinfBD} leads to a larger interval $[\omega_1,\omega_2]$, i.e., an expanded searching space. Note that as predominant acceleration frequency boundary of human-driven vehicles $\omega_1$ tends to $0$, the $\mathcal{H}_{\infty}^{[\omega_1,\omega_2]}$ norm value converges to $1$ (due to $\lim_{\omega \to 0^{+}} |F(j\omega)| = 1$, and based on definitions \eqref{SCon} and \eqref{HinfBD}), i.e., the same value for the $\mathcal{H}_{\infty}$ norm value.

\subsection{Case 2: a large communication delay}

\begin{figure}[!ht]
    \centering
    \includegraphics[scale = 0.22]{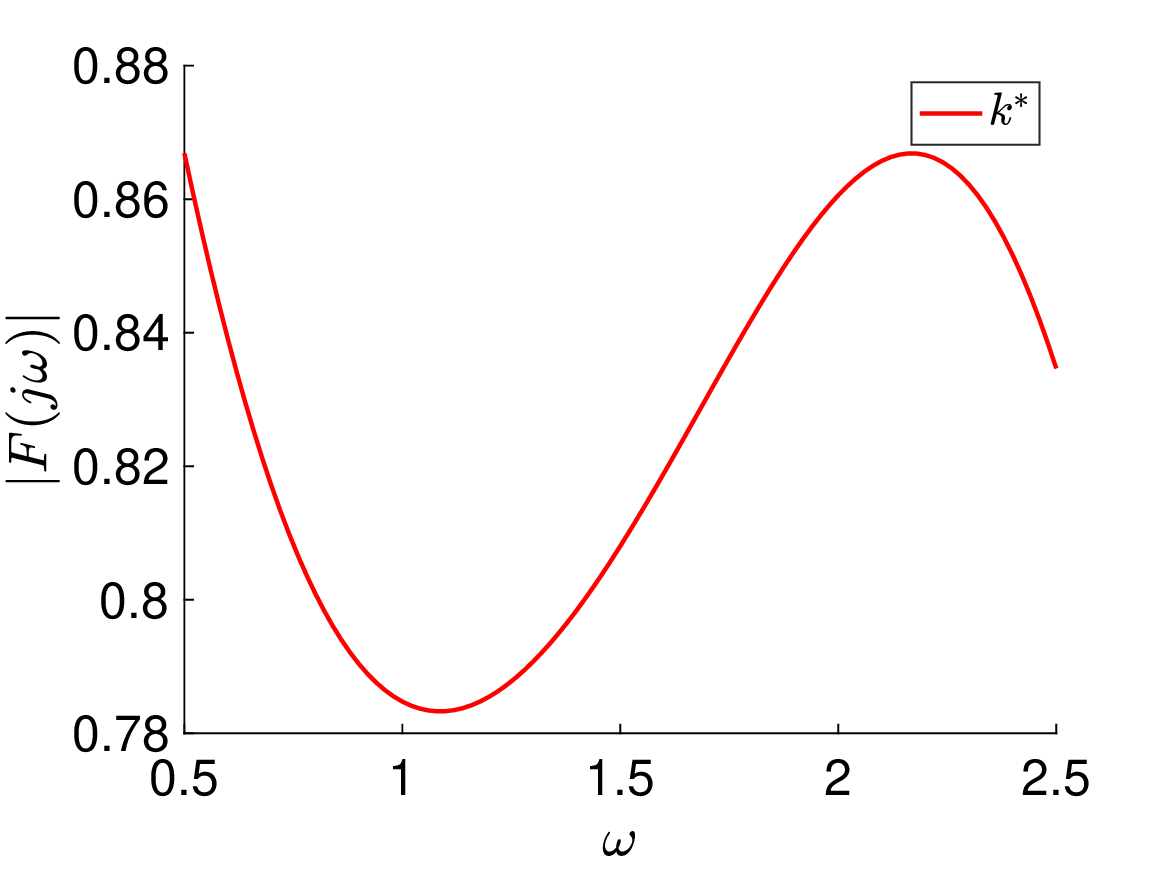}
    \includegraphics[scale = 0.22]{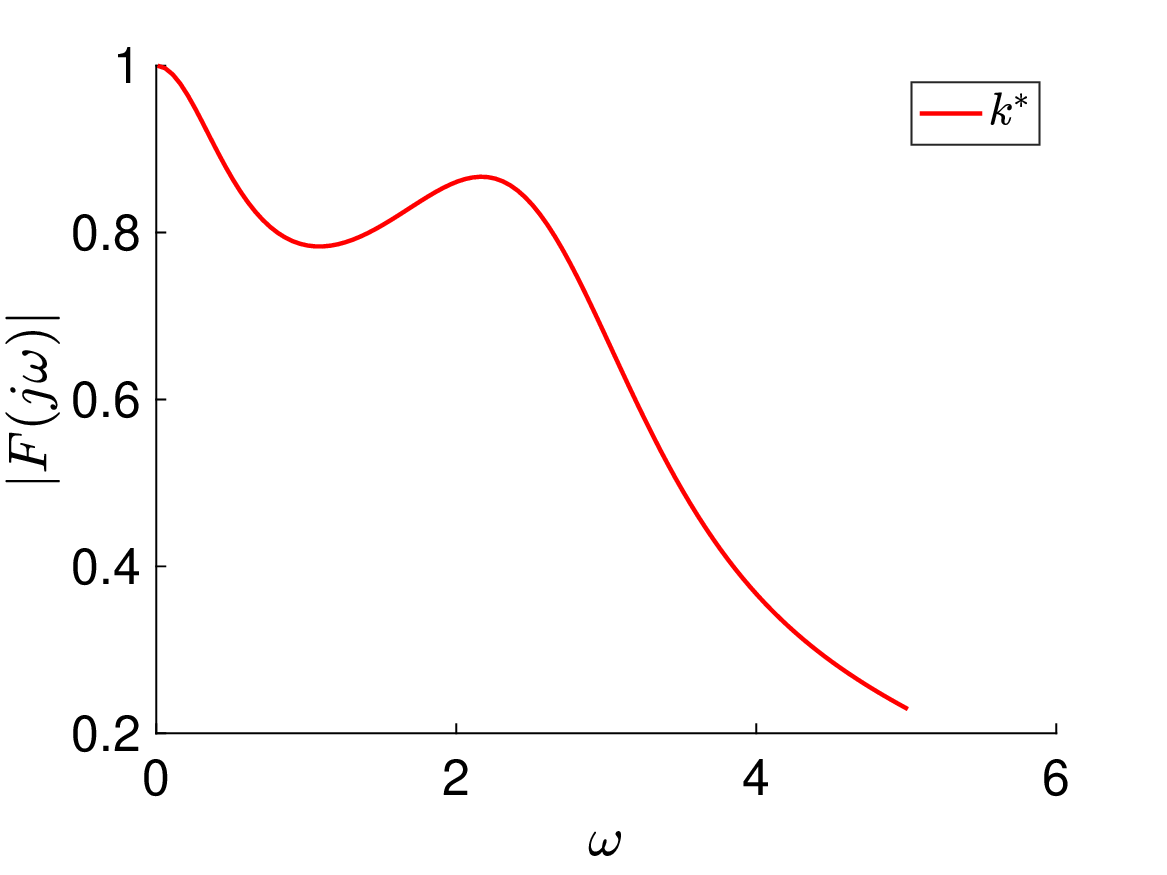}
    
    \caption{The $\omega$-dependent $|F(j\omega)|$ values corresponding to the constrained $\mathcal{H}_{\infty}$ controller synthesis $k^{\ast}$ for $\omega \in [\omega_1,\omega_2]$ on the left and $\omega \in [0.01,5.01]$ on the right.}
    \label{fig3}
\end{figure}

In this section, we corroborate the usefulness of the two-stage $\mathcal{H}_{\infty}$ controller synthesis procedure presented by Procedure \ref{proc:TSA} in the case of dealing with large $\theta$. Repeating the experiments for $\theta = 1.5$, we obtain the two-stage $\mathcal{H}_{\infty}$ controller synthesis illustrated by Tab. \ref{tab:my_label4} for which the $\mathcal{H}_{\infty}^{[\omega_1,\omega_2]}$ norm value corresponding to $k^{\ast}$ is $\|F(s)\|_{\infty}^{[\omega_1,\omega_2]} = 0.8669$. Accordingly, Fig. \ref{fig3} visualizes the $\omega$-dependent $|F(j\omega)|$ values corresponding to the constrained $\mathcal{H}_{\infty}$ controller synthesis $k^{\ast}$ for $\omega \in [\omega_1,\omega_2]$ on the left and $\omega \in [0.01,5.01]$ on the right. According to Tab. \ref{tab:my_label4}, $\lambda(A + Bk^{\ast}) = \{-3.2621, -0.2376 - 0.3644j, -0.2376 + 0.3644j \}$, and Fig. \ref{fig3}, we observe that box constraints \eqref{BC} and both local stability \eqref{LSC} and string stability \eqref{SCon} are satisfied for $k^{\ast}$. 

\begin{table}[t]
    \centering
    \caption{The two-stage $\mathcal{H}_{\infty}$ controller synthesis for the value setting in Tab. \ref{tab:my_label} except for $\theta$ with $\theta = 1.5$, $\omega_1 = 0.5$, $k^{l} = - 2\begin{bmatrix}
    0 & 1 & 1 & 1
\end{bmatrix}^\top$, $k^{u} = 2\mathbf{1}$, $\alpha = 1.05$, $\zeta = 5$, $\nu = 5$, $N = 5$.}
\label{tab:my_label4}
    \begin{tabular}{|c|c|}
        \hline
        Vector & Value \\
        \hline
        $k^{\ast}$ & $\begin{bmatrix}
    1.9696 & 1.9953 & -0.2273 & 0.0234
\end{bmatrix}^\top$\\
        \hline
        $\kappa^{\ast}$ & $\begin{bmatrix}
    0.8341 & 1.3187 & -0.1138 & -0.0214
\end{bmatrix}^\top$\\
        \hline
        $\kappa^0$ & $\begin{bmatrix}
    -0.2638 & 0.5087 & 0.1669 & -0.3410
\end{bmatrix}^\top$\\
        \hline
        $k^0$ & $\begin{bmatrix}
    0.4219 & 1.8308 & -1.1174 & 0.3717
\end{bmatrix}^\top$\\
        \hline
        $\mu^0$ & $\begin{bmatrix}
    -0.8649 & -0.0940 & 0.8405 & 0.3706
\end{bmatrix}^\top$\\
        \hline
    \end{tabular}
\end{table}

\section{Concluding Remarks} \label{Con}

To effectively address the limitations arising from the unconstrained CAV platoon controller synthesis in the literature, this work proposes a (near-optimal) constrained CAV platoon controller synthesis subject to the lower and upper bounds on the control parameters. To that end, given box constraints on the control parameters, parameterizing the set of feedback and feedforward gains to ensure the local stability of the CAV platoon and deriving a necessary condition to ensure the string stability criterion additionally, we obtain more representative parameterized locally stabilizing feedback and feedforward gains. Then, we employ the Padé approximation to more accurately ensure the string stability criterion in the presence of communication delay. 

We empirically observe that such an approximation outperforms the Taylor series approximation in the literature. Due to the string unstable behavior of human-driven vehicles, we have no direct control over them. However, applying the (fully) CAV platoon controller synthesis results to the mixed vehicular platoon can effectively attenuate the stop-and-go disturbance amplification throughout the mixed vehicular platoon. Considering the mixed vehicular platoon scenario and minimizing the $\mathcal{H}_{\infty}$ norm of the Padé approximated transfer function over an interval defined by predominant acceleration frequency boundaries of human-driven vehicles, we solve for a near-optimal constrained CAV platoon controller synthesis via non-convex and non-smooth optimization tools. Conducting extensive numerical experiments corroborates that \textit{(i)} in the case of a sufficiently small communication delay, the Padé approximation-based method outperforms the Taylor series approximation in terms of the $\mathcal{H}_{\infty}$ norm over an interval defined by predominant acceleration frequency boundaries of human-driven vehicles and \textit{(ii)} in the case of a large communication delay, the Padé approximation-based method successfully proposes an $\mathcal{H}_{\infty}$ controller synthesis while the Taylor series approximation-based method in the literature becomes unusable. 

Modern transportation systems will potentially benefit from the proposed CAV control strategy in terms of effective stop-and-go disturbance attenuation as it will potentially reduce collisions. Furthermore, taking advantage of the sensory data, data-driven CAV platoon controllers can be synthesized to enhance the disturbance attenuation performance. Specifically, in the case of inaccurate models, such data-driven counterparts will show their potential superiority over model-based synthesis in terms of disturbance attenuation performance.

\textit{Limitations}: Although the near-optimal performance of the proposed controller synthesis is satisfactory, its near-optimality level can be sensitive to the first stage (stable initialization) of the proposed two-stage $\mathcal{H}_{\infty}$ controller synthesis procedure. It is noteworthy that enhancing the quality of the non-convex and non-smooth optimization tools also affects the quality of the proposed controller synthesis in terms of the near-optimality level. As another limitation, in the current study, we have assumed that the vehicle dynamics and the communication delay are deterministic and time-invariant, which is unrealistic. Such a limitation necessitates the consideration of the uncertain and time-varying nature of those elements \cite{zhou2019robust,zhou2020stabilizing} to develop a robust and adaptive CAV platoon controller synthesis. Elaboration on such a generalized synthesis is left as a pertinent future direction.

\bibliographystyle{IEEEtran}
\bibliography{References}

\appendices

\section{Proof of Proposition \ref{Propo1}} \label{App1}

    Imposing box constraints \eqref{BC1}--\eqref{BC3} to the parameterization \eqref{PZ}, we obtain
    \begin{subequations} \label{IB}
    \begin{align} 
        &k_1^l \le x \le k_1^u,\\
        &k_2^l \le -\tau x + y \le k_2^u,\\
        &k_3^l \le \frac{-Tx+y}{Ky}-z \le k_3^u. 
    \end{align}
    \end{subequations}
    Since $0 < x$, $0 < y$, and $0 < z$ hold, we can equivalently consider them as $\epsilon \le x$, $\epsilon \le y$, and $\epsilon \le z$, respectively, for an infinitesimal $\epsilon > 0$. Then, \eqref{IB} along with $\epsilon \le x$, $\epsilon \le y$, $\epsilon \le z$, and $K > 0$ implies that
    \begin{subequations} \label{IEC}
        \begin{align}
            \epsilon &\le x,\\
            k_1^l &\le x,\\
            x &\le k_1^u,\\
            \epsilon &\le y,\\
            \tau x + k_2^l &\le y,\\
            \frac{Tx}{-Kk_3^l + 1 -K \epsilon} &\le y, \label{IIB}\\
            y &\le \tau x + k_2^u,\\
            \epsilon &\le z, \label{IC1} \\
            \frac{-Tx+y}{Ky}-k_3^u &\le z,\\
            z &\le \frac{-Tx+y}{Ky}-k_3^l, \label{IC2}
        \end{align}
    \end{subequations}hold. Notice that \eqref{IIB} is obtained from the combination of \eqref{IC1} and \eqref{IC2}, and then simplifying the resulting inequality $\epsilon \le \frac{-Tx+y}{Ky}-k_3^l$ via $K > 0$. Thus, utilizing \eqref{IEC}, the $x$, $y$, and $z$ in \eqref{PZ} satisfying box constraints \eqref{BC1}--\eqref{BC3} can be parameterized as \eqref{xyz} and the proof is complete.

    \section{Proof of Lemma \ref{Lem1}} \label{App2}

Utilizing the derivative rules in classic Calculus, it can be verified that
\begin{align*}
    & \frac{d |F_i(j \omega)|}{d \omega} = \frac{D(\omega)N'(\omega)-D'(\omega)N(\omega)}{2D(\omega)\sqrt{D(\omega)N(\omega)}},
\end{align*}
holds. Since 
\begin{align*}
    & \lim_{\omega \to 0^+} N(\omega) = n_0,~\lim_{\omega \to 0^+} D(\omega) = d_0,\\
    & \lim_{\omega \to 0^+} N'(\omega) = 0,~\lim_{\omega \to 0^+} D'(\omega) = 0,
\end{align*}
are all satisfied, we obtain 
\begin{align} \label{DRes}
    \lim_{\omega \to 0^+} \frac{d |F_i(j \omega)|}{d \omega} &= 0.
\end{align}It can similarly be verified that (drop $\omega$ for the brevity)
\begin{align*}
    & \frac{d^2 |F_i(j \omega)|}{d \omega ^2}=\\ & \frac{D^2(2NN''-(N')^2)+ N^2(3(D')^2-2DD'')-2DND'N'}{4D^2N\sqrt{DN}},
\end{align*}
holds. Since 
\begin{align*}
    & \lim_{\omega \to 0^+} N(\omega) = n_0,~\lim_{\omega \to 0^+} D(\omega) = d_0,\\
    & \lim_{\omega \to 0^+} N'(\omega) = 0,~\lim_{\omega \to 0^+} D'(\omega) = 0,\\
    & \lim_{\omega \to 0^+} N''(\omega) = 2(n_2 + \lim_{\omega \to 0^+} g_{\theta}(\omega)),~\lim_{\omega \to 0^+} D''(\omega) = 2d_2,\\
    & \lim_{\omega \to 0^+} g_{\theta}(\omega) = -2K^2 k_4 k_1,~n_0 = d_0 = K^2 k_1^2,
\end{align*} 
are all satisfied, we obtain 
\begin{align} \label{DDRes}
    \lim_{\omega \to 0^+} \frac{d^2 |F_i(j \omega)|}{d \omega ^2} &= \frac{n_2-n_0-2K^2k_4k_1}{n_0} = \frac{-\eta}{K^2 k_1^2},
\end{align}
where $\eta$ denotes the LHS of \eqref{SSC3}. According to \eqref{DRes} and \eqref{DDRes}, if \eqref{SCon} (i.e., a sufficient condition to guarantee the string stability) holds for $F_i(s)$ in \eqref{TF}, then $\lim_{\omega \to 0^+} \frac{d^2 |F_i(j \omega)|}{d \omega^2}  = \frac{-\eta}{K^2 k_1^2} \le 0$ is satisfied (proof by contradiction incorporating the second-derivative test \cite{thomas1992calculus}) and consequently, $\eta \ge 0$ holds. Thus, inequality \eqref{SSC3} holds for the control parameters $\begin{bmatrix}
    k_1 & k_2 & k_3 & k_4
\end{bmatrix}$ and the proof is complete. 

\section{Proof of Proposition \ref{Prop2}} \label{App3}

Imposing box constraints \eqref{BC} to the parameterizations \eqref{PZ} and \eqref{k4p}, we obtain
    \begin{subequations} \label{IBB}
    \begin{align} 
        &k_1^l \le x \le k_1^u,\\
        &k_2^l \le -\tau x + y \le k_2^u,\\
        &k_3^l \le \frac{-Tx+y}{Ky}-z \le k_3^u,\\
        & k_4^l \le \frac{\tau^2 x}{2} - \tau y + \frac{Tx}{Ky} + z + w \le k_4^u.
    \end{align}
    \end{subequations}
    Since $0 < x$, $0 < y$, and $0 < z$ hold, we can equivalently consider them as $\epsilon \le x$, $\epsilon \le y$, and $\epsilon \le z$, respectively, for an infinitesimal $\epsilon > 0$. Then, \eqref{IBB} along with $\epsilon \le x$, $\epsilon \le y$, $\epsilon \le z$, $0 \le w$, and $K > 0$ implies that
    \begin{subequations} \label{IECC}
        \begin{align}
            & \epsilon \le x,\\
            & k_1^l \le x,\\
            & x \le k_1^u,\\
            & \epsilon \le y,\\
            & \tau x + k_2^l \le y,\\
            & \frac{Tx}{-Kk_3^l + 1 -K \epsilon} \le y, \label{IIBB}\\
            & \frac{\xi(x) + \sqrt{\xi(x)^2 + \frac{4T\tau x}{K}}}{2 \tau} \le y, \label{YY} \\
            & y \le \tau x + k_2^u,\\
            & \epsilon \le z, \label{IC11} \\
            & \frac{-Tx+y}{Ky}-k_3^u \le z,\\
            & z \le \frac{-Tx+y}{Ky}-k_3^l, \label{IC22}\\
            & z \le \frac{-\tau^2 x}{2} + \tau y + \frac{-Tx}{Ky}+ k_4^u, \label{ZZ}\\
            & 0 \le w, \label{W1}\\
            & \frac{-\tau^2 x}{2} + \tau y + \frac{-Tx}{Ky} - z + k_4^l \le w,\\
            & w \le \frac{-\tau^2 x}{2} + \tau y + \frac{-Tx}{Ky} - z + k_4^u, \label{W2}
        \end{align}
    \end{subequations}hold. Notice that \eqref{IIBB} is obtained from the combination of \eqref{IC11} and \eqref{IC22}. Similarly, \eqref{ZZ} is obtained from the combination of \eqref{W1} and \eqref{W2}. Furthermore, \eqref{YY} is obtained from the combination of \eqref{IC11} and \eqref{ZZ}, and then equivalently imposing the non-negativity of the quadratic polynomial $\mathcal{P}(y) = \tau y^2 - \xi(x) y - \frac{Tx}{K}$ (as $0 < y$ holds) with $\xi(x) = \frac{\tau^2 x}{2} + \epsilon - k_4^u$. Notice that for such a quadratic polynomial, $\Delta = \xi(x)^2 + \frac{4T\tau x}{K} > 0$ holds, and since $\big(-\frac{Tx}{K}\big)/\tau = -\frac{Tx}{K\tau} < 0$ is satisfied, the polynomial has a positive real root (i.e., the larger root) $0 < r_{+} = \frac{\xi(x) + \sqrt{\xi(x)^2 + \frac{4T\tau x}{K}}}{2 \tau}$ and a negative real root (i.e., the smaller root) $r_{-} = \frac{\xi(x) - \sqrt{\xi(x)^2 + \frac{4T\tau x}{K}}}{2 \tau} < 0$. For $y \in \mathbb{R}$, we know that $0 \le \mathcal{P}(y)$ holds if and only if $y \le r_{-}$ or $r_{+} \le y$ holds. However, $y \le r_{-}$ is not feasible as $0 < y$ and $r_{-} < 0$ hold. Thus, $0 \le \mathcal{P}(y)$ holds if and only if $r_{+} \le y$, i.e., \eqref{YY} holds. Thus, utilizing \eqref{IECC}, the $x$, $y$, $z$, and $w$ in \eqref{PZ} and \eqref{k4p} satisfying box constraints \eqref{BC} can be parameterized as \eqref{wpar} and the proof is complete.

\end{document}